\documentclass[aps,nofootinbib,twocolumn,prd,eqsecnum,showpacs,showkeys,preprintnumbers]{revtex4-1}
\usepackage[caption=false]{subfig}
\usepackage{graphicx}
\usepackage{amsmath}
\usepackage{amsfonts}
\usepackage{amssymb}
\usepackage{color}
\usepackage{bm}
\usepackage{mathrsfs}
\usepackage{epstopdf}
\usepackage{url}
\usepackage{footnote}
\usepackage{textcomp}
\usepackage{ulem}
\usepackage{esint}
\usepackage[unicode=true, pdfusetitle,
 bookmarks=true,bookmarksnumbered=false,bookmarksopen=false,
 breaklinks=false,pdfborder={0 0 1},backref=false,colorlinks=false]{hyperref}
\usepackage{multirow}
\usepackage{pifont}
\begin{document}
\title{Scalar warm inflation in holographic cosmology}
\author{Zahra Bouabdallaoui$^{1}$}
\email{zahraandto@hotmail.com}
\author{Ahmed Errahmani$^{1}$}
\email{ahmederrahmani1@yahoo.fr}
\author{Mariam Bouhmadi-L\'{o}pez$^{2,3}$}
\email{mariam.bouhmadi@ehu.es}
\author{Taoufik Ouali$^{1}$}
\email{ouali1962@gmail.com}
\affiliation{$^{1}$Laboratory of Physics of Matter and Radiation, \\
University of Mohammed first, BP 717, Oujda, Morocco\\
$^{2}$Department of Physics University of the Basque Country UPV/EHU. P.O. Box 644, 48080 Bilbao, Spain\\
$^{3}$IKERBASQUE, Basque Foundation for Science, 48011, Bilbao, Spain}

\date{\today}

\begin{abstract}
We consider warm inflation in the context of holographic cosmology. The weak and the strong dissipative regime are analysed in the slow-roll approximation and within what is known as intermediate inflation. For an appropriate choice of the equation of state, the intermediate inflation is not only an exact solution in general relativity, but also in the holographic setup considered in this paper. Within this approach several dissipative  and physically relevant  functions are considered. We constrain our model using the latest Planck data. We conclude that three of the models analysed are consistent with Planck data for some ranges of the model parameters. However, one of them is ruled out by the observations.
\end{abstract}

\maketitle
\section{Introduction}
 The inflationary paradigm was proposed to sort out several shortcomings  of the standard Big Bang theory. \cite{1, Albrecht, Starobinsky, Linde2, Sato, Linde1}. In addition, this paradigm reproduces correctly the distribution of the large scale structure \cite{6, 7, 8, 9, 10} and the observed anisotropy of the cosmic microwave background \cite{Bennett, 11, 12, 2, Larson}. The zoo of the inflationary models is quite wild and many models are consistent with the current observations. Therefore, it would be quite interesting to have some approach that alleviate such a degeneracy \cite{Arxi17, Arxi18, Arxi19, Arxi20, Arxi21}.  Within the inflationary scenario with a slow roll approach, two roads can be taken. On the one hand, in the cold inflationary scenario, the inflaton does not interact with its environment and the reheating of the Universe takes place at the end of inflation when the scalar field reach the bottom of the potential and starts oscillating. However, this kind of approach might face some fine-tunning problems and the reheating problem itself. Furthermore, the latest Planck data ruled out many forms of inflaton potentials, including, for instance, the quadratic and quartic potentials, both well motivated from a particle physics perspective. On the other hand, there is the warm inflationary (WI) scenario, where the radiation production occurs gradually as the Universe inflates through the dissipation of the inflaton field into a thermal radiation \cite{18,3}. 
In addition, initial density fluctuations (the seeds of the large scale structure) is produced during the WI era i.e. the fluctuation of the inflaton field is due to  a thermal process rather than to a quantum one as in cold inflation \cite{4, 28, 29, 30, 31}. This thermal process affect deeply the background and the perturbative results of the inflationary dynamics \cite{Ber, Bast}.\par

Furthermore, cold and warm inflation have been well studied in the framework of an effective approach to  quantum gravity. Indeed, for example in extra-dimensions models, motivated by superstring theory \cite{Stewart:1994ts, Dvali:1998pa}  such as brane-world models where matter is considered to be  confined in a 3-brane while gravity can propagate in the bulk, have been a natural arena for inflation. In addition, warm inflation has been analysed in modified theory of gravity, like brane-worlds  \cite{delCampo:2007cia, Cid:2007fk, Deshamukhya:2009wc, Nozari:2009th, Romero:2009mg, BasteroGil:2011mr, Herrera:2014nta, Herrera:2015aja, Kamali:2019xnt, Cid} and in loop quantum cosmology \cite{Herrera11,Zhang_2013,Herrera_2014,Graef,Benetti_2019}.
We would to remember that Randall and Sundrum model \cite{Randall:1999vf} is extremely well motivated and can be used as the main setup of the AdS/CFT correspondence. We recall that the AdS/CFT correspondence, conjectured by Maldacena \cite{Maldacena}, consists in describing the 5-gravitational theory as a conformal field theory on the 4-D boundary space-time. This duality can be seen as an holographic approach. The effect of the holographic picture on cold inflation has been analysed in \cite{Lidsey, ZahraPRD} by a conformal field theory on its 4-dimensional boundary space-time. Previous study of the effect of this holographic picture on cold inflation scenario might be found in Refs. \cite{Lidsey, ZahraPRD}, and its extension by a non minimal coupling (NMC) to the induced gravity by means of the dynamical system and the modifications of the amplitude of the primordial perturbation in \cite{AatifaNPB, AatifaIJMPD}.  Recently, we have studied a NMC Higgs inflation in this context \cite{AatifaHiggs}.\\

In this paper, we study warm inflation within an holographic view point in order to complete our previous study \cite{ZahraPRD} and to see how well is warm inflation supported within the holographic picture. In this context, we consider a well known form of the scale factor and named as intermediate inflation. This kind of model was already studied in general relativity and in a modified theory of gravity \cite{Barrow1, Barrow2, Herrera3, Herrera4, Herrera5, Herrera2}.\\

The outline of the paper is as follow. In Sec. \ref{secII}, we construct the main formulas of warm inflation in the context of the holographic duality. In Sec. \ref{secIII}, we show how the intermediate inflation occurs from an equation according to the holographic setup. In Sec. \ref{secIV}, we review the main formulas used in warm inflation at the perturbative level. In Secs. \ref{secV} and  \ref{secVI} we consider the case of the weak and the strong dissipation regime respectively. In these sections, we compute the main parameters of these regime in terms of the holographic parameter and of the perturbation parameters. We conclude our paper in section \ref{secVII}.


\section{HOLOGRAPHIC WARM INFLATION}\label{secII}
In the AdS/CFT correspondence, the Friedmann equation is modified as \cite{Kiritsis,Lidsey}
\begin{equation}\label{equation de fridmann1}
H^2=\frac{2\kappa \rho_c}{3}\Big(1+\epsilon\sqrt{1-\frac{\rho}{\rho_c}}\Big),
\end{equation}
where $\kappa=\hat{m}_p^{-2}=8\pi m_p^{-2}$ and   $m_p$  is the 4-dimensional Planck mass, $\rho$ is the total energy density, $\rho_c={3\hat{m}_p^{4}}/{8c}$ and $c={\hat{m}_p^{6}}/{M^6}$ is the conformal anomaly coefficient which is defined as the ratio between  the  4-dimensional Planck mass and the 5-dimensional mass $M$. At the low-energy limit, $\rho \ll \rho_c$, and for the branch $\epsilon=-1$ the standard form of the Friedmann equation can be recovered. In the following only this branch will be considered.\\
In warm inflation, we will consider that the universe is filled with a self-interacting scalar field with energy density $\rho_{\phi}$ and a radiation energy density $\rho_{\gamma}$.\\

The conservation law of the total energy density $\rho=\rho_{\phi}+\rho_{\gamma}$ reads
\begin{equation}\label{equation de conservation}
\dot{\rho}_{\phi}+3H(\rho_{\phi}+p_{\phi})+\dot{\rho}_{\gamma}+4H\rho_{\gamma}=0
\end{equation}

The dynamical equations for the energies densities, $\rho_{\phi}$ and $\rho_{\gamma}$, in warm inflation are described respectively by \cite{3}
\begin{equation}\label{equation de conservation1}
\dot{\rho}_{\phi}+3H(\rho_{\phi}+p_{\phi})=-\Upsilon\dot{\phi}^2,
\end{equation}
\begin{equation}\label{equation de conservation2}
\dot{\rho}_{\gamma}+4H\rho_{\gamma}=\Upsilon\dot{\phi}^2,
\end{equation}
where a dot means  derivative with respect to the cosmic time. The positive dissipation factor, $\Upsilon$, is responsible for the reheating the universe through the decay of the scalar field $\phi$. Several phenomenological expression for the dissipation term $\Upsilon\dot{\phi}^2$ are given in the literature \cite{32,Gleiser,33, 34, Zhang, Bastero, Bastero1}.\\

Following Refs.\cite{Zhang, Bastero}, we consider the general form of the dissipative coefficient, given by
\begin{equation}\label{dissipation factor1}
\Upsilon(\phi,T)=C_{\phi}\frac{T^m}{\phi^{m-1}}
\end{equation}
where $m$ is an integer and $C_{\phi}$ is associated to the dissipative microscopic dynamics. Different expressions for the dissipation coefficient, i.e. different choices of  values of $m$, have been analyzed in \cite{Zhang, Bastero, Bastero1}. In this paper, we will discuss the following cases

\begin{itemize}
\item $m=-1$  which  corresponds to the dissipation rate of the non supersymmetric model \cite{Gleiser,33}.
\item $m = 0$ in which  the dissipation coefficient represents an exponentially decaying propagator in the high temperature regime. A dissipation coefficient which depends only on the scalar field was first considered in warm inflation in \cite{Oliveira}.
\item  $m = 1$  corresponds to the high temperature regime \cite{Panotopoulos,Ber, Moss2,Bastero2}.
 \item $m = 3$  is motivated by a supersymmetric model \cite{Ber, BasteroB, Bastero}, a minimal warm inflation \cite{Berghaus,Laine,Motaharfar}, a quantum field theory model of inflation \cite{Bastero-Gil} and through axion inflation \cite{DeRocco}. Furthermore, this case is used in other contexts such as in a hilltop model \cite{Bueno}, in a stochastic approach \cite{Ramos2}, in a potential with an inflection point \cite{Cerezo} and in runway potentials \cite{Das_2020}.  By using a sphaleron rate in a non-abelian gauge fields \cite {Berghaus}, a Chern-Simons diffusion rate in a minimal warm inflation \cite {Laine} and in the Dirac-Born infeld \cite {Motaharfar}, the authors show that the dissipative coefficient in the case $m = 3$ does not depend on the inflaton fields as in Eq. (\ref{dissipation factor1}).
  \end{itemize}
	
The energy density, $\rho_{\phi}$, and the pressure, $p_{\phi}$, of the standard scalar field can be written as
\begin{equation}\label{pression et densite}
\rho_{\phi}=   \frac{\dot{\phi}^2}{2} +V(\phi) ,\quad \quad p_{\phi}=\frac{\dot{\phi}^2}{2} - V(\phi),
\end{equation}
where $V(\phi)$  represents the effective potential.\\

By introducing the dimensionless dissipation parameter $Q$, defined as
\begin{equation}\label{RGH}
Q\equiv\frac{\Upsilon}{3H},
\end{equation}
Eq. (\ref{equation de conservation1}) can be rewritten as
\begin{equation}\label{equation de conservation3}
\dot{\rho}_{\phi}=-3H\dot{\phi}^2(1+Q).
\end{equation}
We will consider two dissipative regimes on this paper. The strong dissipative one characterised by $Q\gg 1$ and the weak dissipative one in which $Q\ll 1$.

At the epoch of warm inflation it is safe to assume that the energy density of the scalar field is the dominant component of the cosmic fluid ($\rho_{\phi}\gg\rho_{\gamma}$ ) \cite{3, 4, 30}.
Therefore, the effective Friedmann Eq. (\ref{equation de fridmann1}) reduces to
\begin{eqnarray}\label{equation de fridmann2}
H^2&\approx&\frac{2\kappa \rho_c}{3}\Big(1-\sqrt{1-\frac{\rho_\phi}{\rho_c}}\Big)\nonumber \\
&=&\frac{2\kappa \rho_c}{3}\Big(1-\sqrt{1-\frac{\frac{\dot{\phi}^2}{2} +V(\phi)}{\rho_c}}\Big),
\end{eqnarray}

Furthermore, by combining Eq. (\ref{equation de conservation3}) and the derivative of Eq. (\ref{equation de fridmann2}), one can show that
\begin{equation}\label{dotphi1}
\dot{\phi}^2=[\dot{\phi}^2]_{\textrm{std}} G,
\end{equation}
where
\begin{equation}\label{dotphis}
[\dot{\phi}^2]_{\textrm{std}}=-\frac{2\dot{H}}{\kappa(1+Q)},
\end{equation}
and
\begin{equation}\label{G}
G=1-\frac{3H^2}{2\kappa\rho_{c}},
\end{equation}
are the kinetic energy density of the scalar field in standard warm cosmology \cite{RamonHerrera} and the correction term characterising the effect of AdS/CFT correspondence, respectively.\par
We can notice that at the low energy limit, ($\rho \ll \rho_c$), the correction term reduces to one and the standard expression
of the scalar field is recovered.

We suppose that during warm inflation the radiation production is quasi-stable, i.e. $\dot{\rho}_{\gamma}\ll4H\rho_{\gamma}$ and $\dot{\rho}_{\gamma}\ll\Upsilon\dot{\phi}^2$ \cite{3, 4, 30}. The combination of Eqs. (\ref{equation de conservation2}) and (\ref{dotphi1}) yields
\begin{equation}\label{rhogamma1}
\rho_{\gamma}=[\rho_{\gamma}]_{\textrm{std}} G,
\end{equation}
where $[\rho_{\gamma}]_{\textrm{std}}$ is the radiation  energy density in standard warm inflation \cite{RamonHerrera} expressed as
\begin{equation}\label{rhogammas1}
[\rho_{\gamma}]_{\textrm{std}}=-\frac{\Upsilon\dot{H}}{2 \kappa H (1+Q)}.
\end{equation}
 Furthermore, the energy density of radiation is related to the temperature as \cite{Kolb1}
\begin{equation}\label{rhogamma2}
\rho_{\gamma}=C_{\gamma}T^4,
\end{equation}
where $C_{\gamma}=\pi^2g_*/30$ and $g_*$ characterise the number of relativistic degrees of freedom. Substituting Eq. (\ref{rhogamma2}) into Eq.(\ref{rhogamma1}), the temperature of the thermal bath $T$ can be written as
\begin{equation}\label{temperature}
T=[T]_{\textrm{std}} G^{\frac{1}{4}},
\end{equation}
where $[T]_{\textrm{std}}$ is the temperature in standard warm inflation \cite{RamonHerrera}
\begin{equation}\label{temperatures}
[T]_{\textrm{std}}=\Big[-\frac{\Upsilon\dot{H}}{2 \kappa C_{\gamma} H (1+Q)}\Big]^{\frac{1}{4}}.
\end{equation}
Substituting Eq. (\ref{temperature}) in (\ref{dissipation factor1}), we get that
\begin{equation}\label{Gamma}
\Upsilon^{\frac{4-m}{4}}=C_{\phi}  \Big[\frac{1}{2 \kappa C_{\gamma} }\Big]^{\frac{m}{4}}   \phi^{1-m}\Big[-\frac{\dot{H}}{ H }\Big]^{\frac{m}{4}}\Big(1+Q\Big)^{-\frac{m}{4}}G^{\frac{m}{4}}.
\end{equation}
At low energy limit, ($\rho \ll \rho_c$), the standard expression \cite{RamonHerrera} of Eq. (\ref{Gamma}) is recovered.

Furthermore, by solving Eq. (\ref{equation de fridmann2}) for the scalar density and using simultaneously Eq. (\ref{dotphi1}), we find the effective potential
\begin{equation}\label{potentiel1}
V(\phi)=\frac{3H^2}{2\kappa}\Big(1+G\Big)+\frac{\dot{H}}{\kappa (1+Q)}\Big(1+\frac{3}{2}Q\Big)G.
\end{equation}

At low energy limit, Eq.(\ref{potentiel1}) recovers the expression of standard warm inflation \cite{RamonHerrera}.\\

 \section{Intermediate inflation}\label{secIII}
In order to illustrate our purpose,  we will focus on the intermediate scenario of inflation \cite{Barrow2, Barrow1, Herrera3, Herrera4, Herrera5, Bast} in which an exact solution can be obtained. Nevertheless, exact solutions can also be found for power law expansion of the universe \cite{55} and in the de Sitter inflationary scenario \cite{1}.
In the intermediate scenario of inflation, the scale factor obeys the following expression \cite{Barrow2, Barrow1, Herrera3, Herrera4, Herrera5, Bast}
\begin{equation}\label{Hubble parameter}
a(t)= a_I \exp (At^f),
\end{equation}
where the two constants $f$ and $A$ satisfy the conditions $0<f<1$ and $0<A$ respectively.  Intermediate inflation model may be derived from an effective theory at low dimensions of a fundamental string theory. Therefore, the study of the intermediate inflationary model is motivated by string/M-theory. In this scenario, the cosmic expansion evolves slower than the standard de Sitter model, $a(t)\propto \exp(H_I t)$ and faster than power law inflation $(a\propto t^p, p > 1)$.

The intermediate scenario of inflation is still an exact solution in our holographic setup. Indeed, an appropriate choice of the equation of state of the form
\begin{equation}
\rho +p=\gamma \rho _{c}\sqrt{1-\frac{\rho }{\rho _{c}}}\left[ (1-\sqrt{1-%
\frac{\rho }{\rho _{c}}})\right] ^{\lambda},  \label{1}
\end{equation}
for $\lambda>1$ and $\gamma>0$ reproduces this intermediate paradigm. For $\lambda=1$ and $\rho\ll\rho_c$, we recover the standard Friedmann equation and the barotropic equation of state $p=(\gamma-1)\rho$. The equation of state (\ref{1}) implies that while the weak energy conditions \cite{IJMPD} is satisfied, the strong energy condition ($\rho+3p\geq 0$) is violated for a sufficiently small energy density and for $\lambda>1$. The violation of the strong energy condition assure that the acceleration  of the Universe takes place.\par
With the help of Eq. (\ref{equation de fridmann1}) and the conservation of the total energy density
\begin{equation}
\overset{\cdot }{\rho }+3H(\rho +p)=0,  \label{2}
\end{equation}
where $\rho=\rho_\phi+\rho_\gamma$ and $p=p_\phi+p_\gamma$ are the energy density and the pressure respectively, a straightforward calculus leads to the intermediate inflation given by Eq. (\ref{Hubble parameter}).\\

{On the other hand, the dimensionless slow-roll parameters are given by
\begin{equation}\label{epsilon1}
\epsilon\equiv-\frac{\dot{H}}{H^2}=\frac{1-f}{Aft^f}
\end{equation}
\begin{equation}\label{eta}
\eta\equiv -\frac{\ddot{H}}{2H\dot{H}}=\frac{2-f}{2Aft^f}.
\end{equation}

 In order to estimate the scale where the perturbations cross the Hubble radius,  we need to define the end of inflation, i.e.  $\epsilon = 1$. According to the behaviour of the potential (see Eq. (\ref{potentiel1})), a description of how inflation ends in our model requires a deeper analysis which is out of the scope of this paper. However, to illustrate our purpose, we have plotted, in Fig. \ref{ew}  and \ref{es}, the potential $V$ and the first slow roll parameter $\epsilon$ for the couple of parameters ($A$, $f$). The weak dissipative regime is illustrated in Fig. \ref{ew} for ($A=0.01$, $f=0.45$), ($A=0.08$, $f=0.27$) and ($A=0.1$, $f=0.13$) and the strong dissipative regime is illustrated in Fig. \ref{es} for ($A=0.1$, $f=0.2$), ($A=0.1$, $f=0.12$) and ($A=0.1$, $f=0.06$). We notice  from Fig. \ref{ew}, that the slow roll parameter is always less than one and its violation cannot stop inflation, in the decreasing branch of the potential, in the cases ($A=0.01$, $f=0.45$) and ($A=0.1$, $f=0.2$) for weak and strong dissipative regime, respectively. In this case, a modification of the model by adding a new a parameter $\phi_{end}$ may trigger the stop of the intermediate inflation \cite{martin2013}. In the case of ($A=0.1$, $f=0.13$) and ($A=0.1$, $f=0.0.06$), the slow roll parameter is always higher than one and the intermediate inflation may occur between values of the scalar field where the scalar field   reaches the maximum of the potential and the scalar field value at which $\epsilon=1$ (first solution) for weak and strong dissipative regimes, respectively. In the last case, the intermediate inflation may also occur  between values of the scalar field at which $\epsilon=1$ (the second solution) and infinite values of the scalar field.

\begin{figure*}[hbtp]
    \centering
    \begin{tabular}{ccc}
      \includegraphics[width=.45\linewidth]{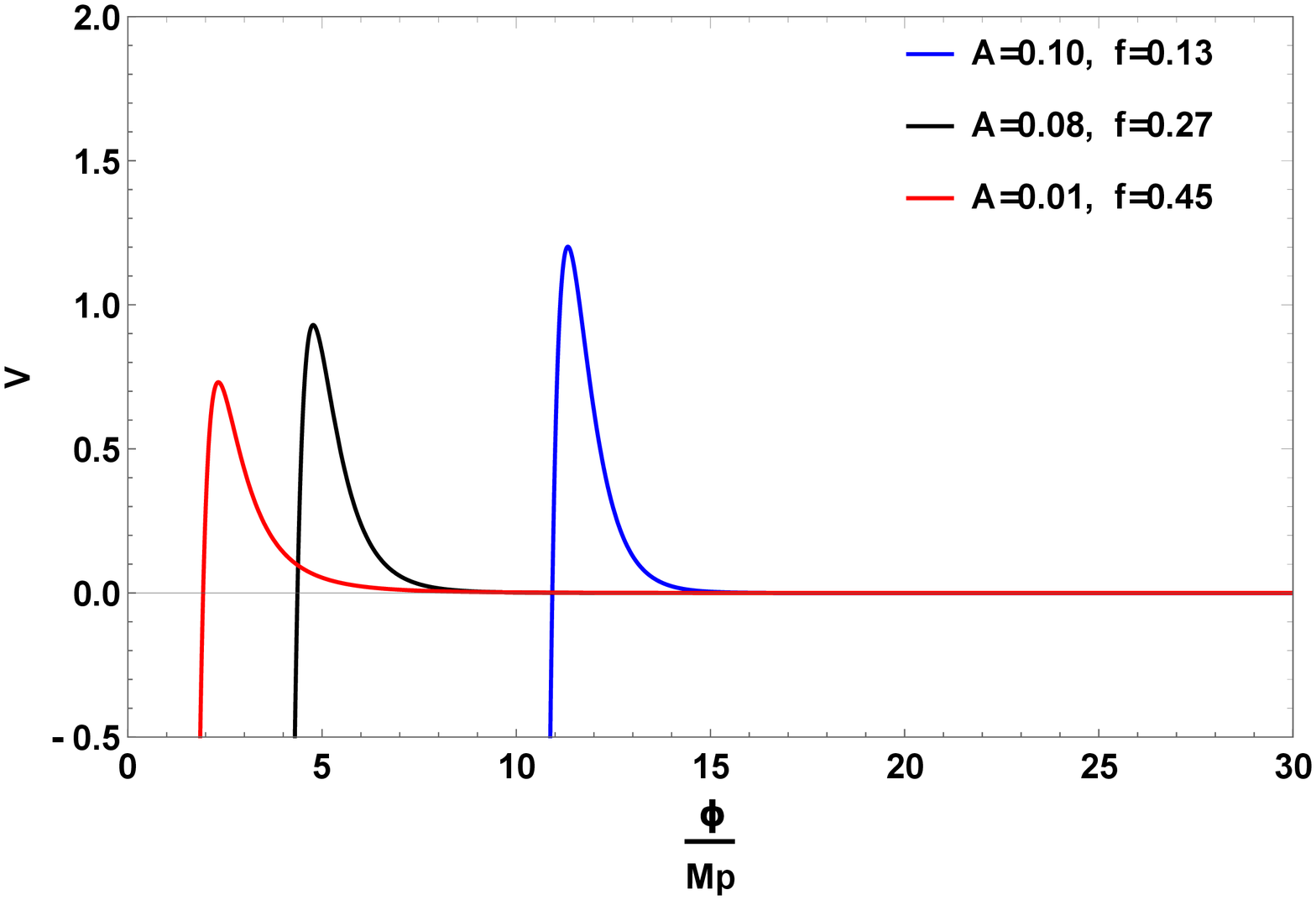}& \qquad &
      \includegraphics[width=.45\linewidth]{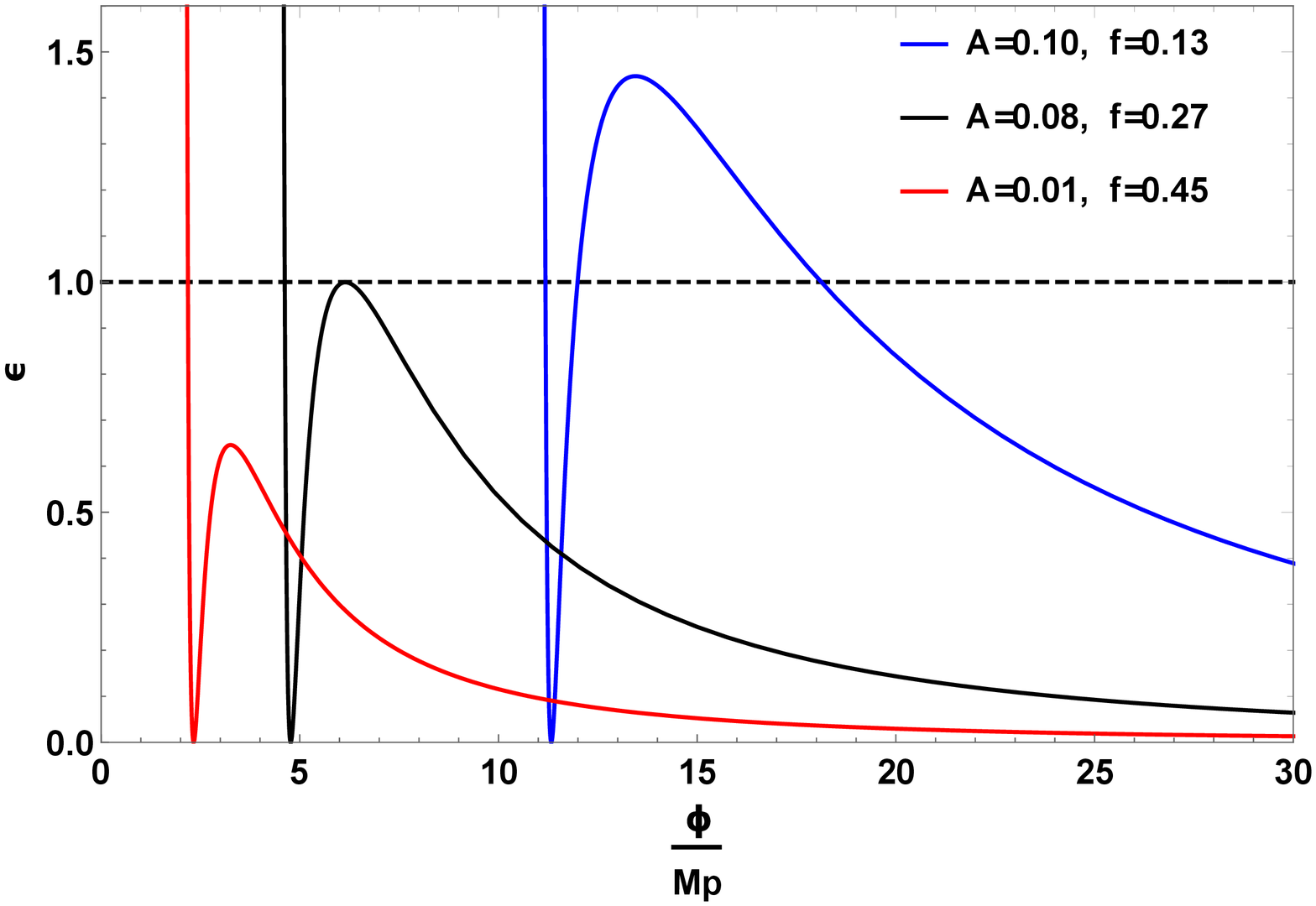}\\
  \end{tabular}
  \caption{Plot of the potential (left curve) and the slow roll parameter (right curve) in the weak dissipative regime for
	($A=0.01$, $f=0.45$), ($A=0.08$, $f=0.27$) and ($A=0.1$, $f=0.13$). The horizontal dashed line (right curve) corresponds to $\epsilon=1$. }\label{ew}
\end{figure*}
\begin{figure*}[hbtp]
    \centering
    \begin{tabular}{ccc}
      \includegraphics[width=.45\linewidth]{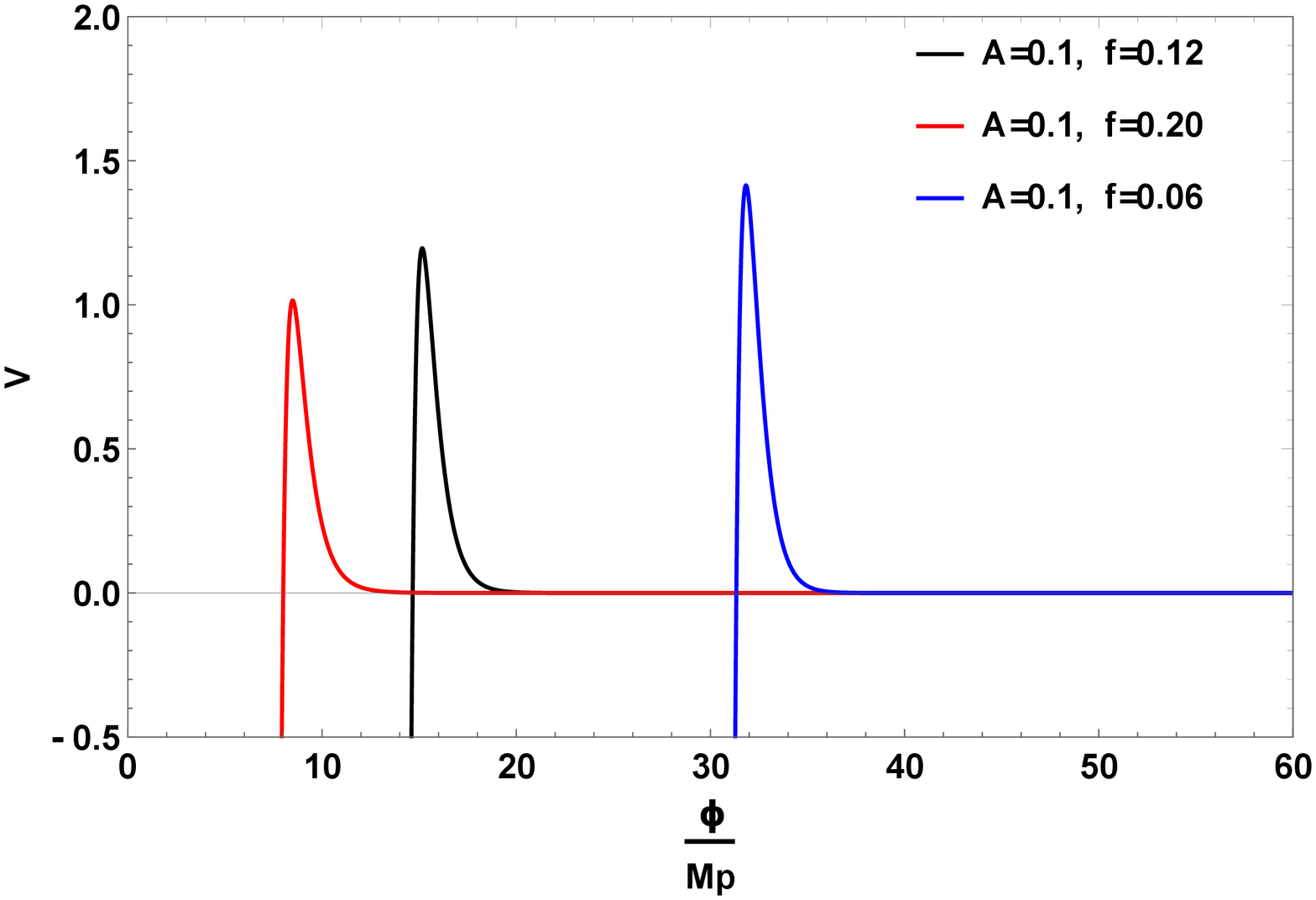}& \qquad &
      \includegraphics[width=.45\linewidth]{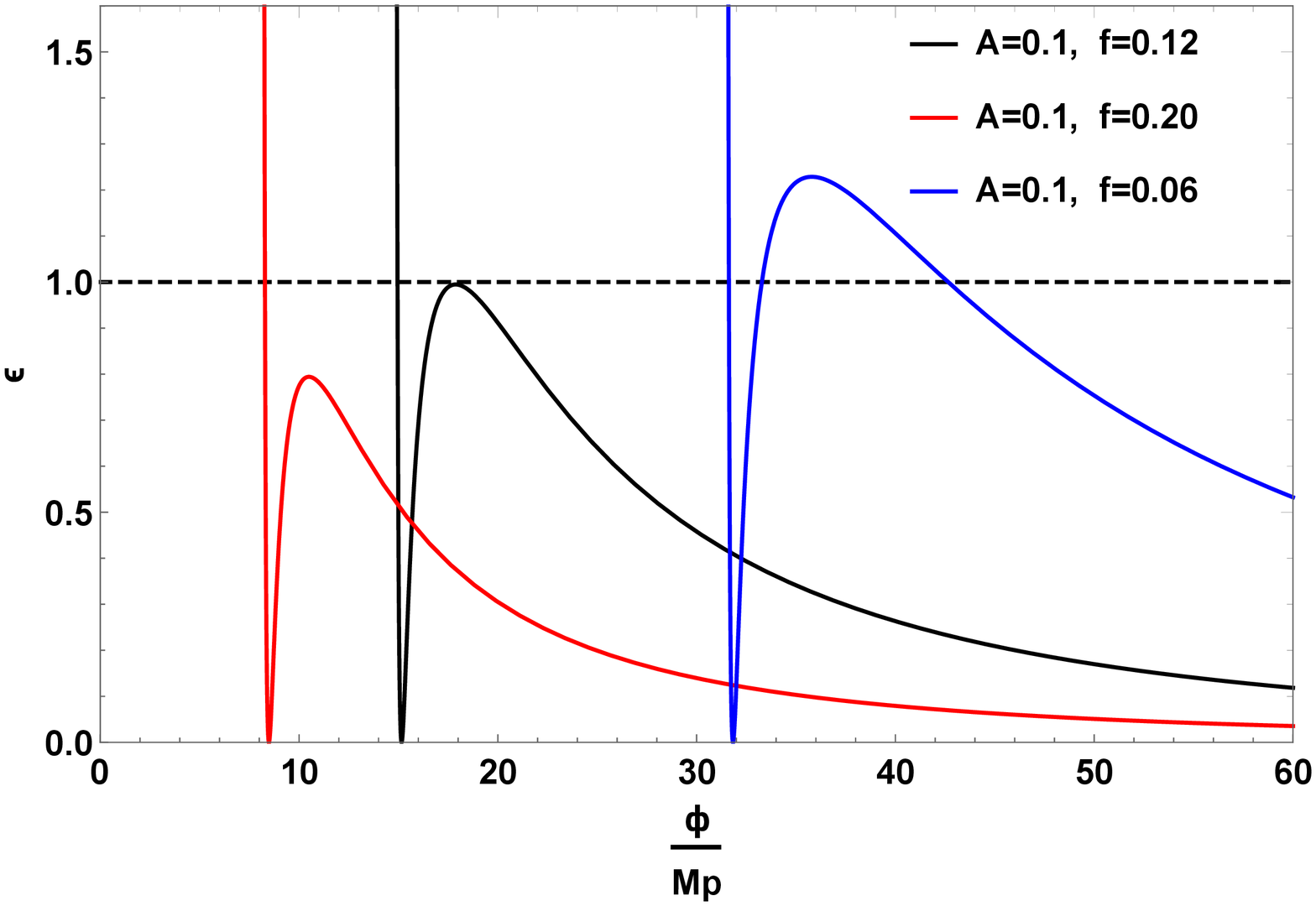}\\
  \end{tabular}
  \caption{Plot of the potential (left curve) and the slow roll parameter (right curve) in the strong dissipative regime for ($A=0.1$, $f=0.2$), ($A=0.1$, $f=0.12$) and ($A=0.1$, $f=0.06$). The horizontal dashed line (right curve) corresponds to $\epsilon=1$.}\label{es}
\end{figure*}

Furthermore, the number e-folds $N$ is given by

\begin{equation}\label{e-folds1}
N=\int^{t_{end}}_{t_k}Hdt=A\Big(t_{end}^f-t_k^f\Big),
\end{equation}
where $t_k$ and $t_{end}$ are the cosmic time at the horizon crossing and at the end of inflation, respectively.
By equating Eq.(\ref{epsilon1}) to unity, we obtain  $t_{end}=\Big( \frac{1-f}{Af}\Big)^\frac{1}{f}$, and  Eq.(\ref{e-folds1}) give us the cosmic time at the horizon crossing
\begin{equation}\label{crossing}
t_{k}\dot{=}I(N)=\Big( \frac{1-f(N+1)}{Af}\Big)^\frac{1}{f}.
\end{equation}  }

{
\section{PERTURBATIONS AND WARM INFLATION} \label{secIV}
Warm inflation  not only affects the background parameters but can also modify those at the perturbative level as the scalar perturbations; though the tensor perturbations remains unchanged.  The general idea for the
perturbations in warm inflation being of thermal origin were given in \cite{3, 4, 28, 29, 30}. However, the most relevant works with the modern notation in which the fluctuations in the scalar field  are mainly originated from thermal fluctuation rather than quantum fluctuations during warm inflation can be found in \cite{Graham_2009,Gleiser,Bastero_Gil_2014,Bastero_2014}.


In order to study the cosmological perturbations, we consider the scalar perturbations of a Friedmann-Lema\^{i}tre-Robertson-Walker (FLRW) background  in the longitudinal gauge. Therefore, the perturbed metric reads
\begin{equation}\label{PM}
ds^2=-(1+2\Phi)dt^2+a^2(t)(1-2\Phi)\delta_{ij}dx^idx^j,
\end{equation}
where $a(t)$ is the scale factor, $\Phi(t,x)$ is the scalar perturbation. Furthermore,
the dissipative coefficient depends explicitly, in general, on the temperature and on the inflaton field as given in Eq (\ref{dissipation factor1}). This dependence modifies the set of the perturbative equations. In fact, the amplitude of the scalar perturbation of the inflaton field will be enhanced by a correction factor. These modifications appear, not only because of the scalar perturbations of the background metric but also due to the existence of a dissipative coefficient in the conservation of the energy momentum tensor through

\begin{equation}
	\delta\Upsilon=\frac{\partial\Upsilon}{\partial T}\delta T+\frac{\partial\Upsilon}{\partial \phi}\delta\phi.
\end{equation}
As we will notice in the next paragraph, this modification will affect strongly the amplitude of the scalar spectrum. This treatment has been already studied in literatures e.g. \cite{Bastero,Bartrum_14,Bastero_Gil_2014,Bastero_2014,Benetti_2017}. In the next paragraph, we introduce the amplitude of the scalar perturbation of the inflaton field as evaluated at the horizon crossing. Like we did previously, we assume that the perturbations will not be highly affected by the holographic approach effect. Therefore, we assume that the amplitude of the scalar  perturbation is given by \cite{Benetti_2017,Bartrum_14,Motaharfar_2019,Das_2020, Bastero_Gil_2014}
\begin{equation}\label{sp}
	P_{{R}}=\left(\frac{H^2}{2\pi\dot{\phi}}\right)^2\left(1+2n+\frac{2\sqrt{3}\pi Q}{\sqrt{3+4\pi Q}}\frac{T}{H}\right)F(Q)
\end{equation}
where the coefficient $n$ denotes the statistical distribution for the inflaton field. By assuming that the inflaton is in thermal equilibrium with the radiation bath, $n$ behaves like a Bose-Einstein distribution. The function $F(Q)$ describes the coupling between the inflaton field and radiation. Its expression is obtained numerically by resolving the systems of perturbations equations for warm inflation \cite{Bastero, Bastero2, Bastero_Gil_2014, Graham_2009}. We will specify its numerical fit in the strong dissipative regime for each value of the dissipative coefficient. From Eqs. (\ref{dotphi1}), (\ref{dotphis}) and (\ref{temperature}), equation (\ref{sp}) can be rewritten as

\begin{widetext}
\begin{equation}\label{warm scalar perturbations3}
P_{R}=\frac{1}{4\pi ^{2}}H^{4}\left[ -\frac{2\dot{H}}{\kappa (1+Q)}\right]
^{-1}G^{-1}\left( 1+2n+\frac{2\sqrt{3}\pi Q}{\sqrt{3+4\pi Q}}\frac{T}{H}%
\right) F(Q),
\end{equation}
where
\begin{equation}\label{T/H}
\frac{T}{H}=\left[ -\frac{\Upsilon\dot{H}}{2\kappa C_{\gamma }H(1+Q)}%
\right] ^{\frac{1}{4}}H^{-1}G^{\frac{1}{4}},
\end{equation}
and the combination of Eqs. (\ref{RGH}) and (\ref{Gamma}) give
\begin{equation}\label{R}
Q\left[ 1+Q\right] ^{\frac{m}{4-m}}=\frac{1}{3H}C_{\phi }^{\frac{4}{4-m}}%
\left[ \frac{1}{2\kappa C_{\gamma }}\right] ^{\frac{m}{4-m}}\phi ^{4\frac{1-m%
}{4-m}}\left[ -\frac{\dot{H}}{H}\right] ^{\frac{m}{4-m}}G^{\frac{m}{4-m}},%
\end{equation}
\end{widetext}
which can be rewritten in term of the parameters of the intermediate inflation, Eq. (\ref{Hubble parameter}), as
\begin{widetext}
\begin{equation}\label{PR}
	P_{R}=\frac{A^{4}f^{4}}{4\pi ^{2}}\left[ \frac{2Af(1-f)}{\kappa (1+Q)}%
\right] ^{-1}I(N)^{3f-2}(1-S(I(N))^{2f-2})^{-1}\left( 1+2n+\frac{2\sqrt{3
}\pi Q}{\sqrt{3+4\pi Q}}\frac{T}{H}\right) F(Q),
\end{equation}
\begin{equation}\label{TH1}
\frac{T}{H}=\frac{1}{Af}\left[ \frac{3Af(1-f)}{2\kappa C_{\gamma }}\frac{Q}{1+Q}\right]
^{\frac{1}{4}}I(N)^{\frac{2-3f}{4}%
}(1-S(I(N))^{2f-2})^{\frac{1}{4}},
\end{equation}
\begin{equation}\label{R1}
Q\left[ 1+Q\right] ^{\frac{m}{4-m}}=\frac{C_{\phi }^{\frac{4}{4-m}}}{3Af}%
\left[ \frac{(1-f)}{2\kappa C_{\gamma }}\right] ^{\frac{m}{4-m}}\phi ^{4%
\frac{1-m}{4-m}}I(N)^{-f+\frac{2m-4}{m-4}}\left[ 1-S(I(N))^{2f-2}\right] ^{%
\frac{m}{4-m}}.%
\end{equation}
\end{widetext}
 where the quantities $S=3A^2f^2/2\kappa\rho_c$ and $I(N)$  are given by Eq. (\ref{crossing}).\\

In the next section, we will examine separately the weak dissipative regime and the strong one. In the weak dissipative case, we will approximate the function $F(Q)$ to be equal to one while for the strong dissipative regime we will fix this function according to the values of the parameters $m$ found in \cite{Benetti_2017, Motaharfar_2019}
\begin{eqnarray}
	F(Q)&\sim & 1+4.981Q^{1.946}+0.127Q^{4.330}\label{m3},\\
	F(Q)&\sim & 1+0.335Q^{1.364}+0.0185Q^{2.315}\label{m1},\\
	F(Q)&\sim & \frac{1+0.4Q^{0.77}}{(1+0.15Q^{1.09})^2}\label{m-1},\\
	F(Q)&\sim 1 \label{m0},
\end{eqnarray}
 for $m=3$, $m=1$, $m=-1$ and $m=0$, respectively.  We can notice, from Eqs. (\ref{m3})-(\ref{m0}),  that in the strong dissipative regime, the scalar perturbations of the inflaton field will be enhanced for the case $m\neq 0$ as compared with the weak dissipative regime.

To complete the set of parameters required to constrain our model by the observational data, we introduce the amplitude of the tensor perturbation, $P_T$.  As mentioned above, while the amplitude of the scalar perturbation in warm inflation is given by Eq. (\ref{sp}), the amplitude of the tensor perturbation remain unchanged and is given by \cite{Motaharfar_2019}
\begin{equation}\label{tensor fluctuations1}
P_T= 2\kappa\Big( \frac{H}{\pi}\Big)^2,
\end{equation}
and from Eq. (\ref{sp}), we can express the tensor-to-scalar ratio, $r={P_{T}}/{P_{R}}$, in terms of the number of e-folds N as
\begin{widetext}
\begin{equation}\label{ratio}
r=\frac{8\kappa }{A^{2}f^{2}}\left[ \frac{2Af(1-f)}{%
\kappa (1+Q)}\right] I(N)^{-f}(1-S(I(N))^{2f-2})\left( 1+2n+\frac{2\sqrt{3}%
\pi Q}{\sqrt{3+4\pi Q}}\frac{T}{H}\right) ^{-1}F(Q)^{-1}.
\end{equation}
\end{widetext}
}
\section{WEAK DISSIPATIVE REGIME } \label{secV}
Considering that the system evolves according to the weak dissipative regime ($\Upsilon \ll 3H$).
Performing the integration of Eq. (\ref{dotphi1}),  the evolution of the scalar field  is given by
\begin{eqnarray} \label{phi1}
\phi_{t}-\phi_{0}&=&\sqrt{\frac{8A(1-f)}{\kappa f}} t^{\frac{f}{2}} H_{A,f,c}(t) \nonumber\\
&=& [\phi_{t}]_{\textrm{std}} H_{A,f,c}(t)
\end{eqnarray}
where the integration constant, $\phi_{0}$, will be set to zero in the rest of the paper and $H_{A,f,c}(t)$, a correction term as compared to standard warm inflation \cite{RamonHerrera}, can be expressed as
\begin{equation}\label{hyp1}
H_{A,f,c}(t)={_2F_1} \Big(-\frac{1}{2},\frac{f}{4(f-1)}; \frac{5f-4}{4(f-1)};S t^{2(f-1)}\Big),
\end{equation}
 where ${_2F_1}(a,b;c;d)$ is the hypergeometric function and $S=\frac{3A^2f^2}{2 \kappa \rho_c }$.

From Eqs. (\ref{Hubble parameter}) and (\ref{Gamma}), as $Q=\Upsilon/3H\ll 1$, the dissipation coefficient $\Upsilon$ can be rewritten in terms of the scalar field as
\begin{equation}\label{Gamma3}
\Upsilon=C_{\phi}^{\frac{4}{4-m}}  \Big[\frac{1-f}{2 \kappa C_{\gamma} }\Big]^{\frac{m}{4-m}}   \phi^{\frac{4(1-m)}{4-m}}t^{-\frac{m}{4-m}}\Big(1-St^{2f-2}\Big)^{\frac{m}{4-m}}.
\end{equation}

The evolution of the scalar  field  and of the dissipation coefficient are given respectively, at the horizon crossing $t_{k}$, by
\begin{equation}\label{phitk}
\phi_{t_{k}}=\sqrt{\frac{8A(1-f)}{\kappa f}} \Big(I(N)\Big)^{\frac{f}{2}} H_{A,f,c}(N),
\end{equation}
and
\begin{eqnarray}\label{Gamma33}
\Upsilon&=&C_{\phi}^{\frac{4}{4-m}}  \Big[\frac{1-f}{2 \kappa C_{\gamma} }\Big]^{\frac{m}{4-m}}   \phi_{t_{k}}^{\frac{4(1-m)}{4-m}}\Big(I(N)\Big)^{-\frac{m}{4-m}}\nonumber\\
&\times&\Big(1-S\Big(I(N)\Big)^{2f-2}\Big)^{\frac{m}{4-m}}.
\end{eqnarray}

If we compare the scalar field and the dissipation coefficient of our model at the horizon crossing  with that obtained in the case of standard warm inflation \cite{RamonHerrera} we find, respectively
\begin{equation}\label{phitks}
\phi_{t_{k}}=[\phi_{t_{k}}]_{\textrm{std}}H_{A,f,c}(N),
\end{equation}
and
\begin{equation}\label{Gamma34}
\Upsilon=[\Upsilon]_{\textrm{std}} [\Upsilon]_{c},
\end{equation}
with
\begin{equation}\label{Gamma34}
[\Upsilon]_{\textrm{std}}=C_{\phi}^{\frac{4}{4-m}}\Big[\frac{1-f}{2 \kappa C_{\gamma} }\Big]^{\frac{m}{4-m}} \phi_{\textrm{std}}^{\frac{4(1-m)}{4-m}}\Big(I(N)\Big)^{-\frac{m}{4-m}},
\end{equation}
and
\begin{equation}\label{Gamma34}
[\Upsilon]_{c}=\Big(H_{A,f,c}(N)\Big)^{\frac{4(1-m)}{4-m}}\Big(1-S(I(N))^{2f-2}\Big)^{\frac{m}{4-m}}.
\end{equation}
Here $H_{A,f,c}(N)$ and $[\Upsilon]_{c}$ represent corrections term to the scalar field and to the dissipation coefficient in the weak dissipation regime, respectively.
We can show that the correction term $H_{A,f,c}(N)$ does not deviate too much from 1 i.e. the variation of the scalar field is not strongly affected  by the  AdS/CFT correspondence.\\

\begin{figure*}[hbtp]
    \centering
    \begin{tabular}{ccc}
     \includegraphics[width=.43\linewidth]{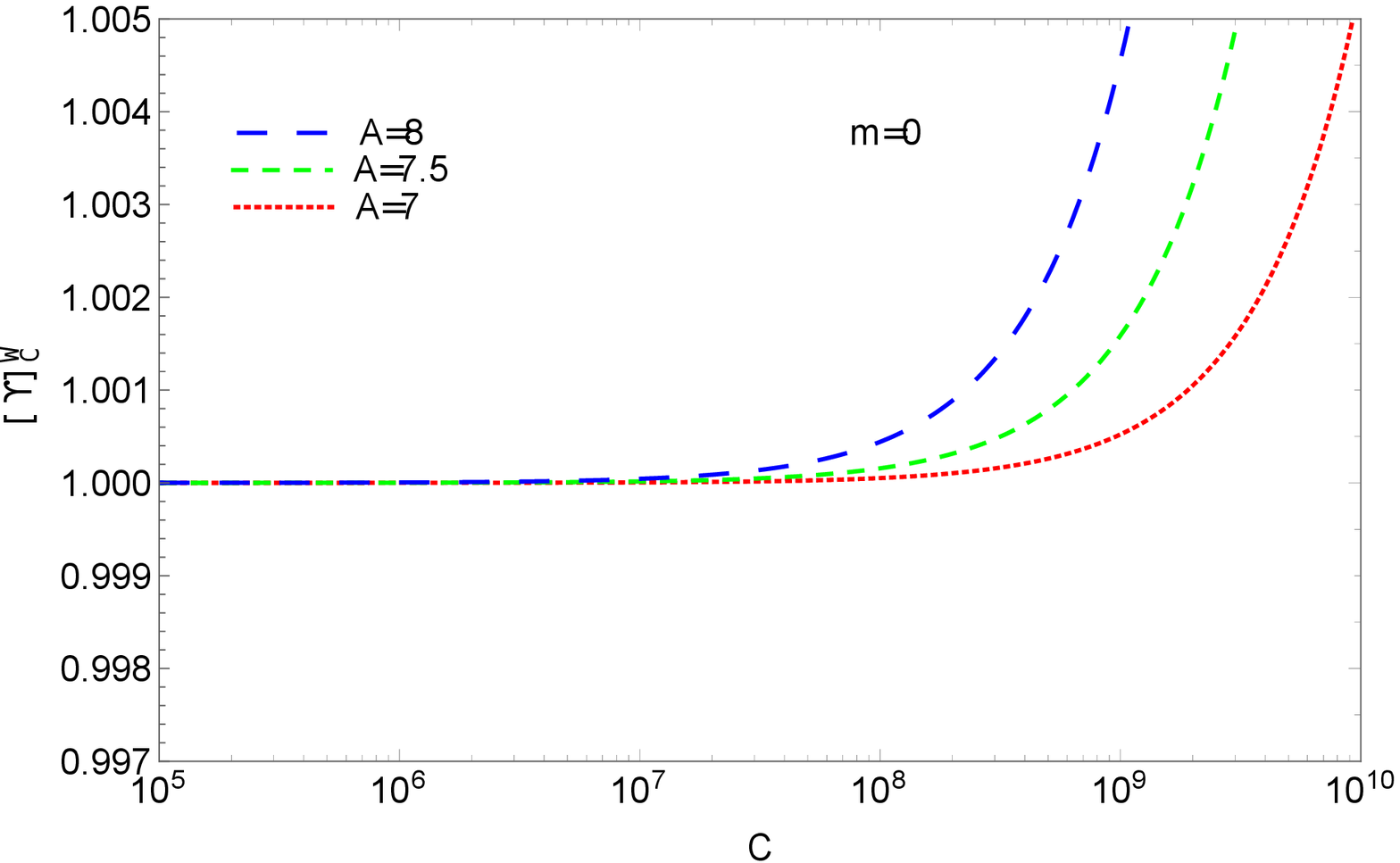} & \qquad\qquad &
     \includegraphics[width=.43\linewidth]{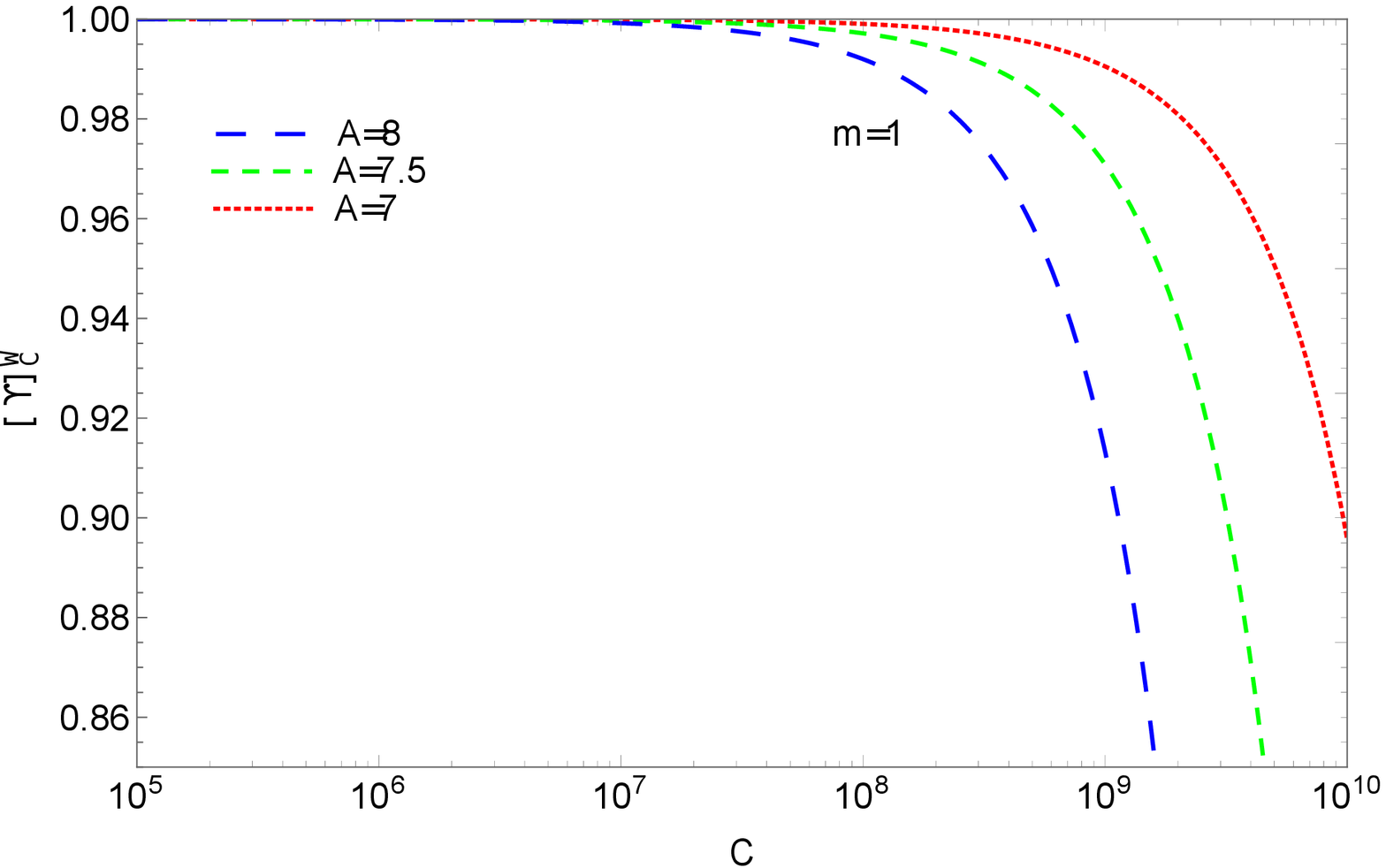} \\

     \includegraphics[width=.43\linewidth]{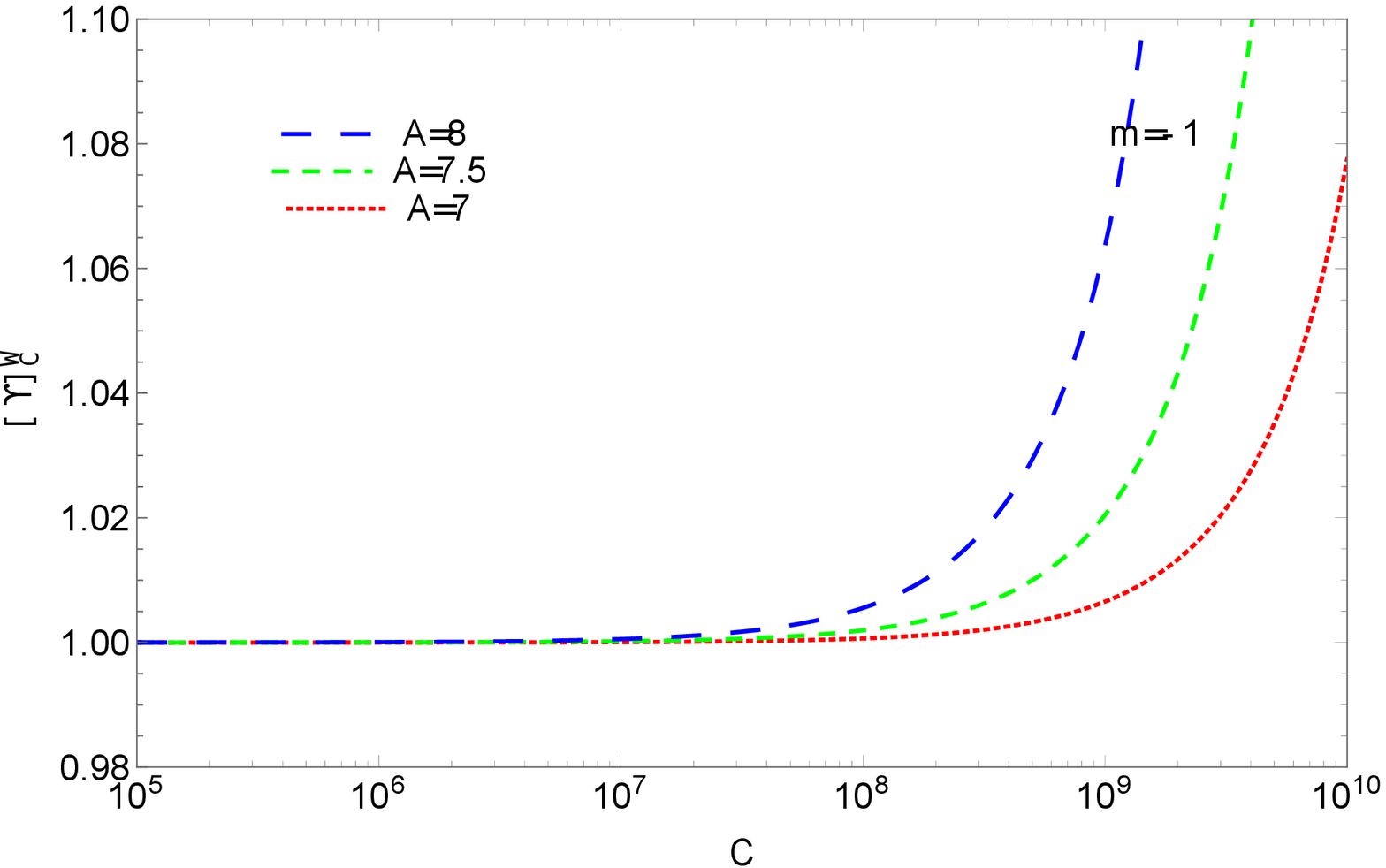} & \qquad\qquad &
     \includegraphics[width=.43\linewidth]{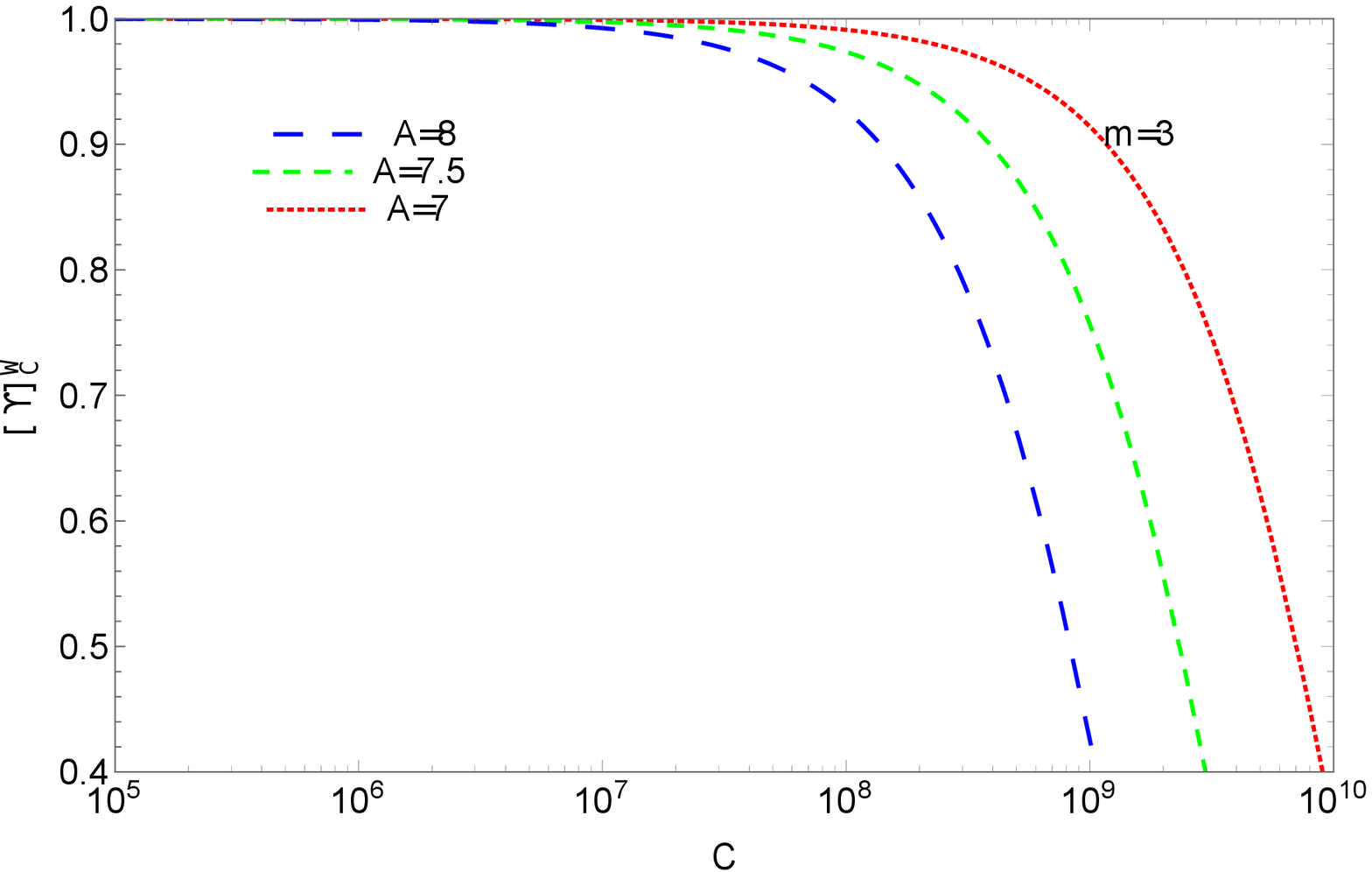} \label{ci}\\
  \end{tabular}
  \caption{Evolution of the correction term $[\Upsilon]^W_c$ versus the conformal anomaly coefficient $c$ in the weak dissipative regime for $f=0.125$ and for three different values of the parameter $A$, for the case $m=0$ (upper left panel), $m=1$ (upper right panel), $m=-1$ (down left panel) and $m=3$ (down right panel). We have used $\kappa=1 $, $N=60$ and $C_{\gamma}=70$.}
\label{fig2}

\end{figure*}

The dissipation term  introduced previously in Eq. (\ref{dissipation factor1}) describes the thermal radiation bath and its evolution versus the conformal anomaly coefficient tell us about the effect of AdS/CFT correspondence on the thermal process. From Fig. \ref{fig2}, we notice that the effect of the AdS/CFT correspondence is more significative and increases with the conformal anomaly coefficient for $c\geq 10^{7}$. This effect increases the thermal process.
 However,  we can see from Eq. (\ref{Gamma33})  that for
specific values of $A$ and $f$ there must be a limiting value $S(I(N))=1$ beyond which the dissipation coefficient no longer makes sense.


\begin{figure*}[hbtp]
    \centering
    \begin{tabular}{ccc}
		(a)   & \qquad\qquad&(b)  \\
\includegraphics[width=.45\linewidth]{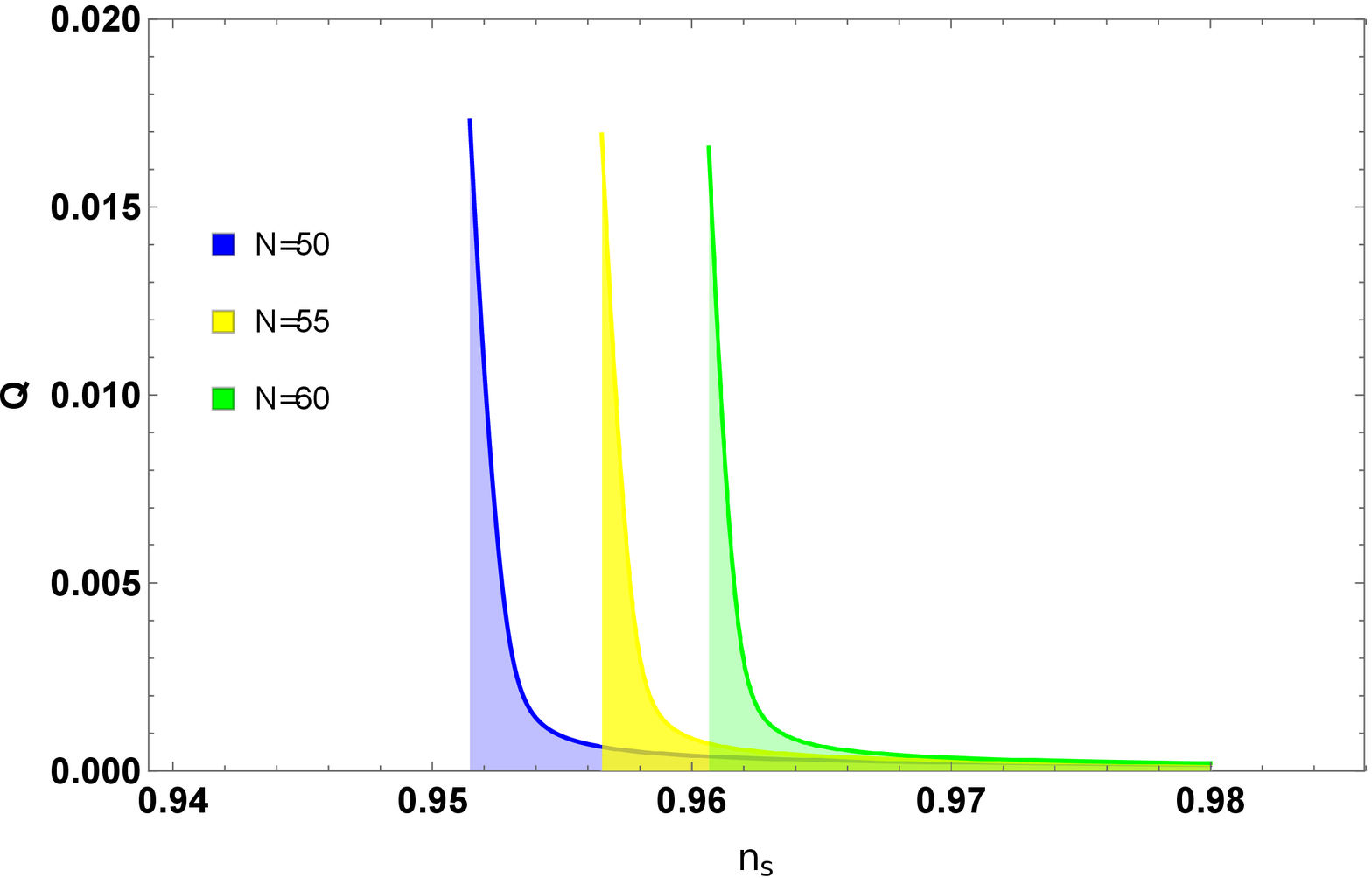}& \qquad\qquad&
     \includegraphics[width=.45\linewidth]{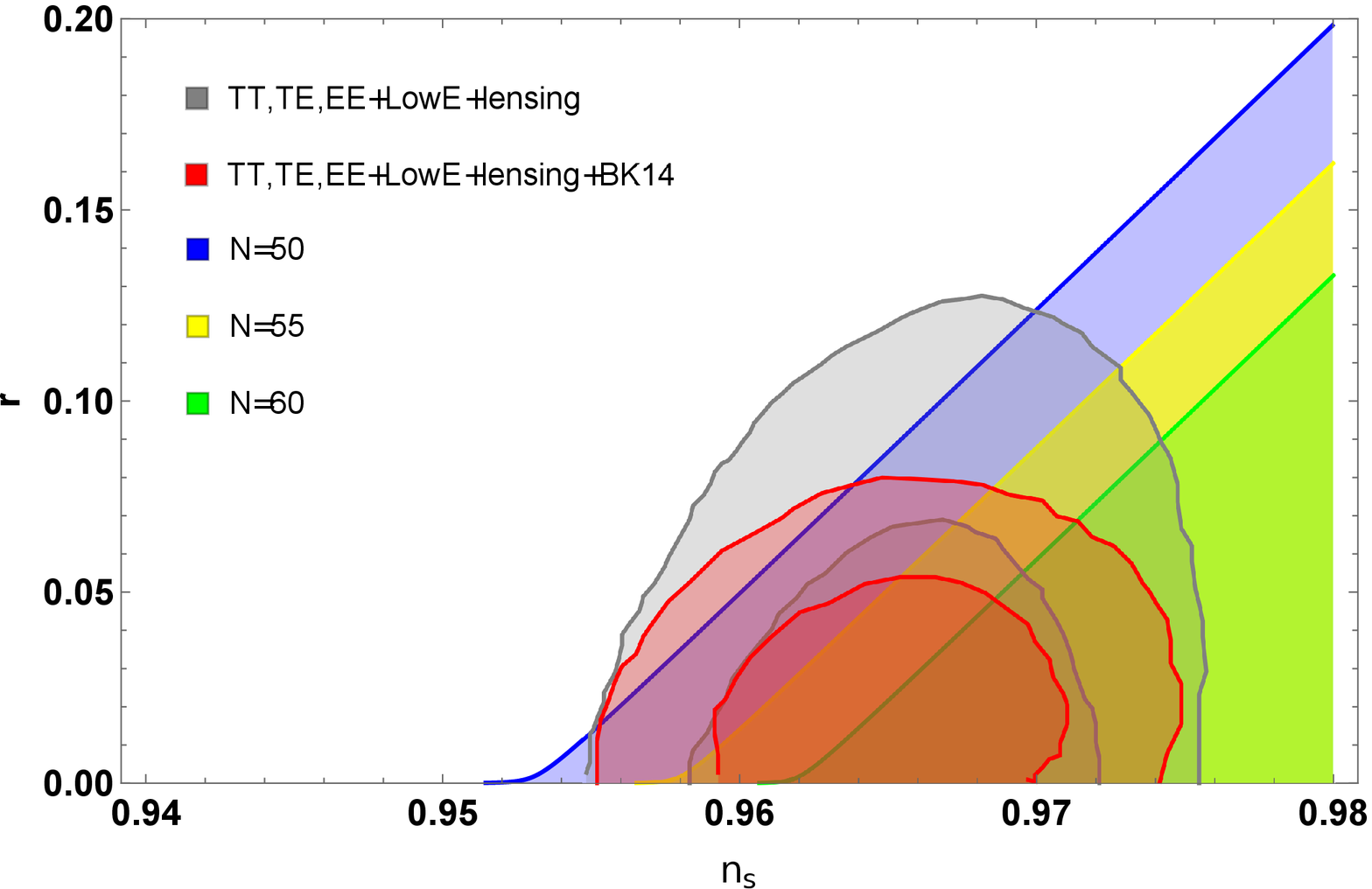}
\end{tabular}
      \caption{The  plots show the variation of the dissipative parameter $Q$ (plot (a)) and the variation of the tensor-to-scalar ratio $r$ (plot (b)) versus the spectral index $n_s$ for  different values of the e-fold number, the conformal anomaly coefficient $c=10^7$ and the parameter $m=-1$.
		These plots are obtained for  the following choice of parameters $C_{\phi }=10^{-25}$, $0.1<A<2$ and $f=0.125$. }
\label{figm=-1}
\end{figure*}

\begin{figure*}[hbtp]
    \centering
    \begin{tabular}{ccc}
    (a)   & \qquad\qquad &(b)  \\
   \includegraphics[width=.45\linewidth]{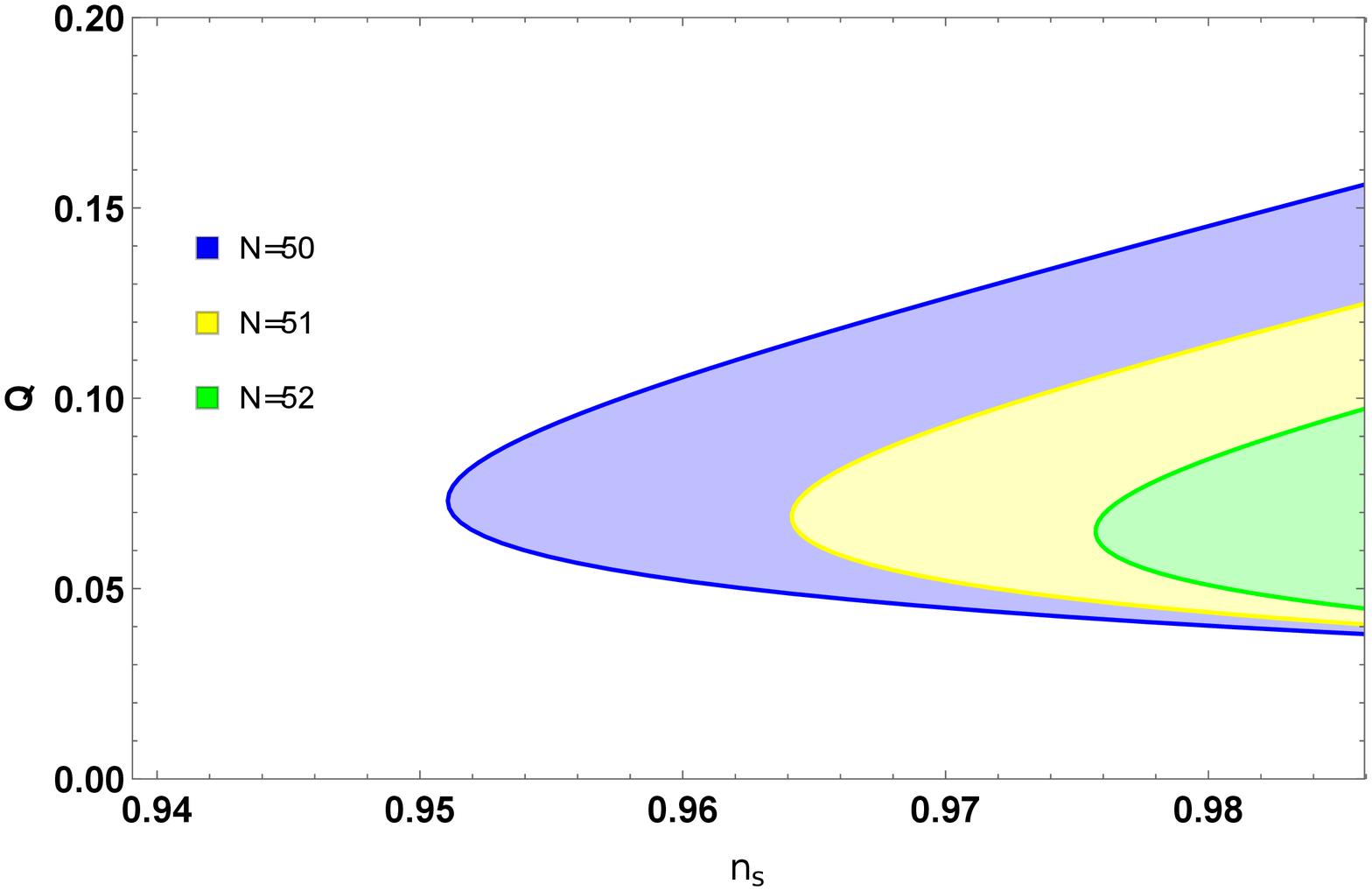}  & \qquad\qquad &  \includegraphics[width=.45\linewidth]{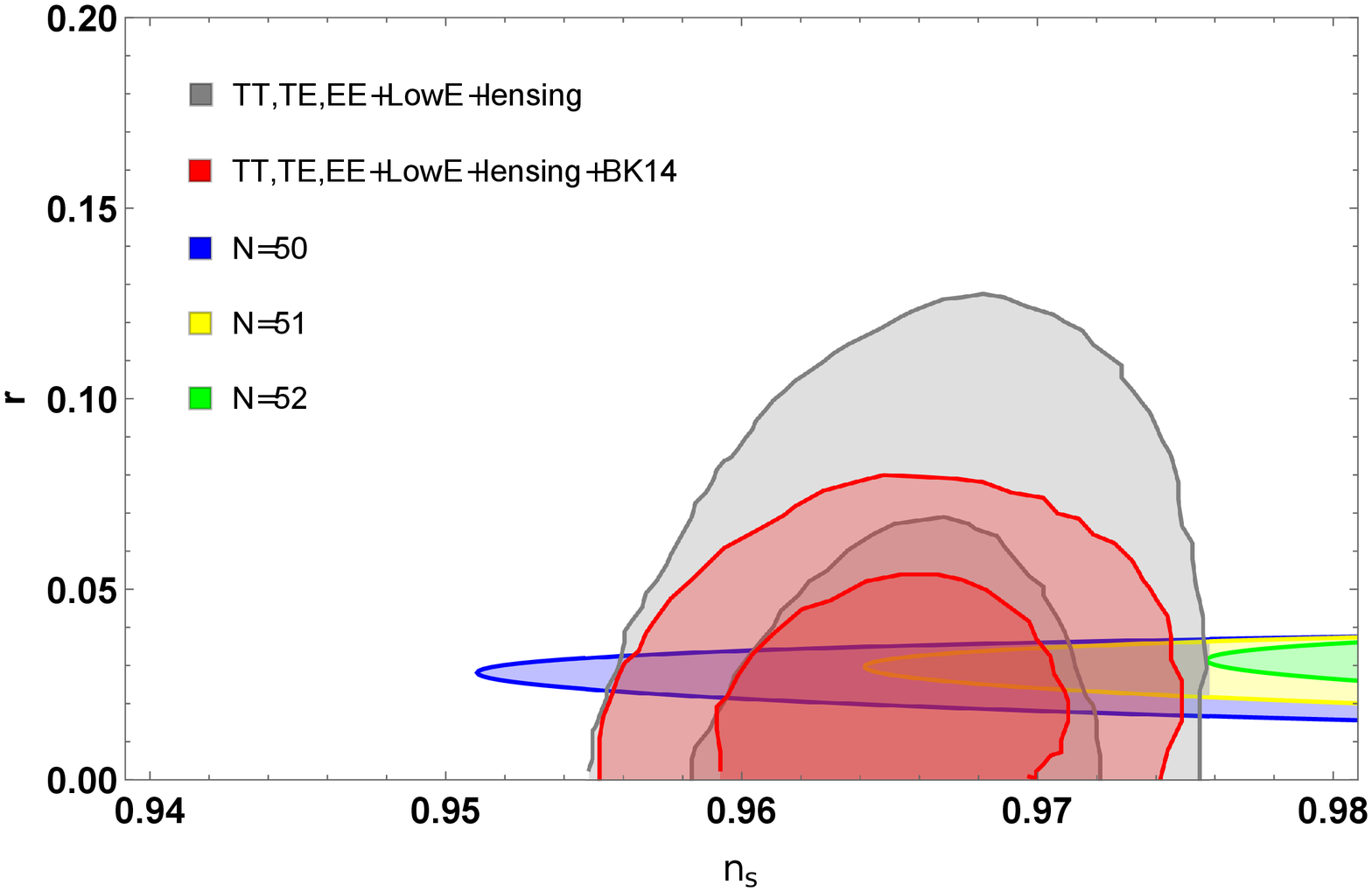} \\
      \end{tabular}
  \caption{The  plots show the variation of the dissipative parameter $Q$ (plot (a)) and the variation of the tensor-to-scalar ratio $r$ (plot (b)) versus the spectral index $n_s$ for different values of the e-fold number, the conformal anomaly coefficient $c=10^7$ and the parameter $m=3$. These plots are obtained for  the following choice of parameters  $C_{\phi }=5.7\times 10^{7}$, $0.1<A<0.5$ and $f=0.5$.} \label{figm=3}
\end{figure*}

{
To incorporate the observed parameters in warm inflation at the perturbative level, we rewrite the amplitude of the scalar perturbation Eqs. (\ref{PR})-(\ref{R1}) and the tensor-to-scalar ratio Eq. (\ref{ratio}) in the weak dissipative regime (i.e. $F(Q)\simeq 1$) as follow
\begin{widetext}
\begin{eqnarray}
P_{R}&=&\frac{A^{4}f^{4}}{4\pi ^{2}}\left[ \frac{2Af(1-f)}{\kappa}
\right]^{-1}I(N)^{3f-2}(1-S(I(N))^{2f-2})^{-1}\left( 1+2n+2\pi Q\frac{T}{H}\right), \label{PR1W}\\
\frac{T}{H}&=&\frac{1}{A.f}\left[ \frac{3Af(1-f)}{2\kappa C_{\gamma }}{Q}\right]
^{\frac{1}{4}}I(N)^{\frac{2-3f}{4}}(1-S(I(N))^{2f-2})^{\frac{1}{4}}, \label{TH1W}\\
Q&=&\frac{C_{\phi }^{\frac{4}{4-m}}}{3Af}\left[ \frac{(1-f)}{2\kappa C_{\gamma }}\right] ^{\frac{m}{4-m}}\phi ^{4\frac{1-m}{4-m}}I(N)^{-f+\frac{2m-4}{m-4}}\left[ 1-S(I(N))^{2f-2}\right] ^{\frac{m}{4-m}},\label{R1W}\\
r&=&\frac{8\kappa }{A^{2}f^{2}}\left[ \frac{2Af(1-f)}{\kappa}\right]
I(N)^{-f}(1-S(I(N))^{2f-2})\left( 1+2n+2\pi Q\frac{T}{H}\right) ^{-1},\label{ratio1W}
\end{eqnarray}
\end{widetext}
where at the crossing horizon time, the inflaton field is given by Eqs. (\ref{phi1}) and (\ref{hyp1}). Furthermore, by combining Eqs. (\ref{TH1W})  and (\ref{R1W}), the above equations read
\begin{widetext}
\begin{eqnarray}
P_{R}&=&\frac{A^{4}f^{4}}{4\pi ^{2}}\left[ \frac{2Af(1-f)}{\kappa}
\right] ^{-1}\text{ }I(N)^{3f-2}(1-S(I(N))^{2f-2})^{-1}\left( 1+2n+2{\pi}
Q\frac{T}{H}\right),\label{PR2W}\\
\frac{T}{H}&=&\frac{1}{Af}\left[ \frac{C_{\phi }(1-f)}{2\kappa C_{\gamma }}
\right] ^{\frac{1}{4-m}}I(N)^{-f+\frac{m-3}{m-4}}\left[ 1-S(I(N))^{2f-2}
\right] ^{\frac{1}{4-m}}\phi ^{\frac{1-m}{4-m}},\label{TH2W}\\
Q&=&\frac{C_{\phi }}{3Af}\left[ \frac{C_{\phi }(1-f)}{2\kappa C_{\gamma }}
\right] ^{\frac{m}{4-m}}\phi ^{4\frac{1-m}{4-m}}I(N)^{-f+\frac{2m-4}{m-4}}
\left[ 1-S(I(N))^{2f-2}\right] ^{\frac{m}{4-m}},\label{R2W}\\
r&=&\frac{8\kappa }{A^{2}f^{2}}\left[ \frac{2Af(1-f)}{\kappa}\right]
I(N)^{-f}(1-S(I(N))^{2f-2})\left( 1+2n+2\pi Q\frac{T}{H}\right) ^{-1},\label{ratio2W}
\end{eqnarray}
\end{widetext}
and the scalar spectral index $n_{s}=1-\frac{d\ln (P_{R})}{dN}$ can be
written as

\begin{widetext}
\begin{eqnarray}
n_{s}-1 &=&
-(3f-2)\frac{d\ln I(N)}{dN}+\frac{d\ln (1-S(I(N))^{2f-2})}{dN}-\frac{d\ln
\left( 1+2n+2\pi Q\frac{T}{H}\right) }{dN}\nonumber\\
&=&-(3f-2)\frac{I(N)^{-f}}{Af}+\frac{2(1-f)SI(N)^{f-2}}{%
Af(1-S(I(N))^{2f-2})}+n_{4},
\end{eqnarray}
\end{widetext}
 where we have used Eq. (\ref{crossing}) and we have defined $n_{4}$ as
\begin{equation}
n_{4}=-\frac{d\ln \left( 1+2n+2\pi Q\frac{T}{H}\right) }{dN}.%
\end{equation}
The  term $n_{4}$ is obtained  numerically. The numerical analysis is done for different values of the number of e-folds and the
conformal anomaly coefficient is set to be $c=10^{7}$. We have also assumed the warm
condition ($\frac{T}{H}>1)$, the weak dissipation case $(Q\ll 1)$ and we have
constrained the spectral index in the observational range $0.9<n_{s}<1$.\\

The results show that the dissipative coefficients $\Upsilon_{1}=C_\phi T$ and $\Upsilon_0=C_\phi \phi$ are not supported by Planck data \cite{12} for any set of values of $C_{\phi }$, $A$ and $f$. However, for a convenient choice of those  parameters, namely $C_{\phi }$, $A$ and $f$, the dissipative coefficients $\Upsilon_{-1}=C_\phi \phi^2/T$  and $\Upsilon_3=C_\phi T^3/\phi^2$ are well supported by data. Indeed,
\begin{itemize}
\item  for $m=-1$, we plot the dissipative parameter $Q$ and the tensor-to-scalar ratio $r$ as
parametric functions of $n_{s}$ for the following range of parameters $C_{\phi }=10^{-25}$, $0.1<A<2$ and $f=0.125$. The parametric plot is shown in Fig. \ref{figm=-1}.
  \item for $m=3$, we plot the dissipative parameter $Q$ and the tensor-to-scalar ratio $r$ as
parametric functions of $n_{s}$ for the following range of parameters $C_{\phi }=5.7\times 10^{7}$, $0.1<A<0.5$ and $f=0.5$. The parametric plot is shown in Fig. \ref{figm=3}.
\end{itemize}
We notice from the plots of Fig. \ref{figm=-1} and \ref{figm=3}, that we are, indeed, in the weak dissipative regime. From the plots of the same Fig. \ref{figm=-1}   i.e. $m=-1$ and $m=3$ in the dissipative coefficient, we notice that the tensor-to-scalar ratio lies in the $1\sigma$ contour of Planck data for $N=55$ and $N=60$. This means that the weak dissipative regime in warm inflation in the holographic context under consideration may reproduce the observational data for the choice of parameters considered above.
}

\begin{figure*}[hbtp]
    \centering
		\begin{tabular}{ccc}
    (a) &\qquad\qquad  &(b) \\
    \includegraphics[width=.47\linewidth]{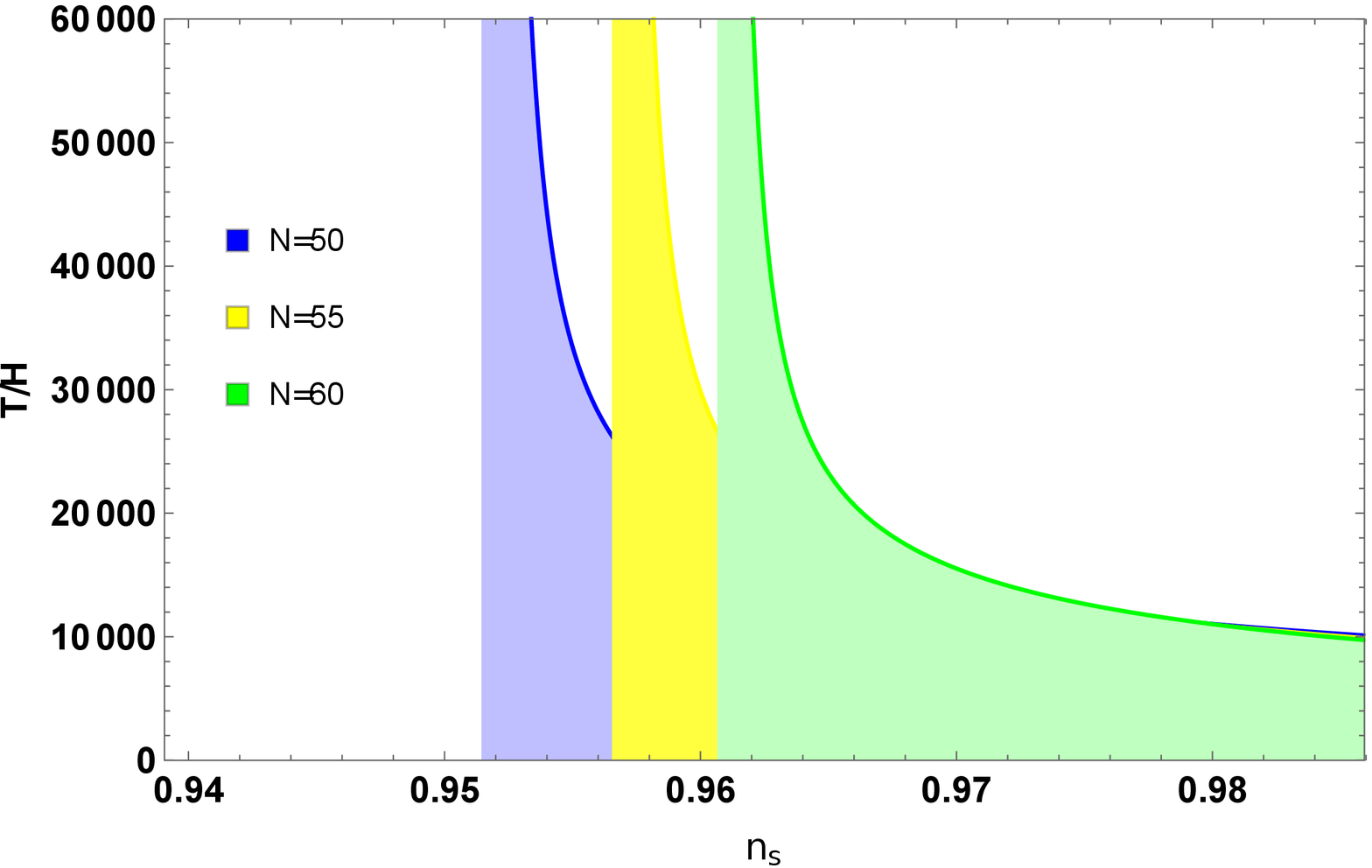}&\qquad\qquad & \includegraphics[width=.45\linewidth]{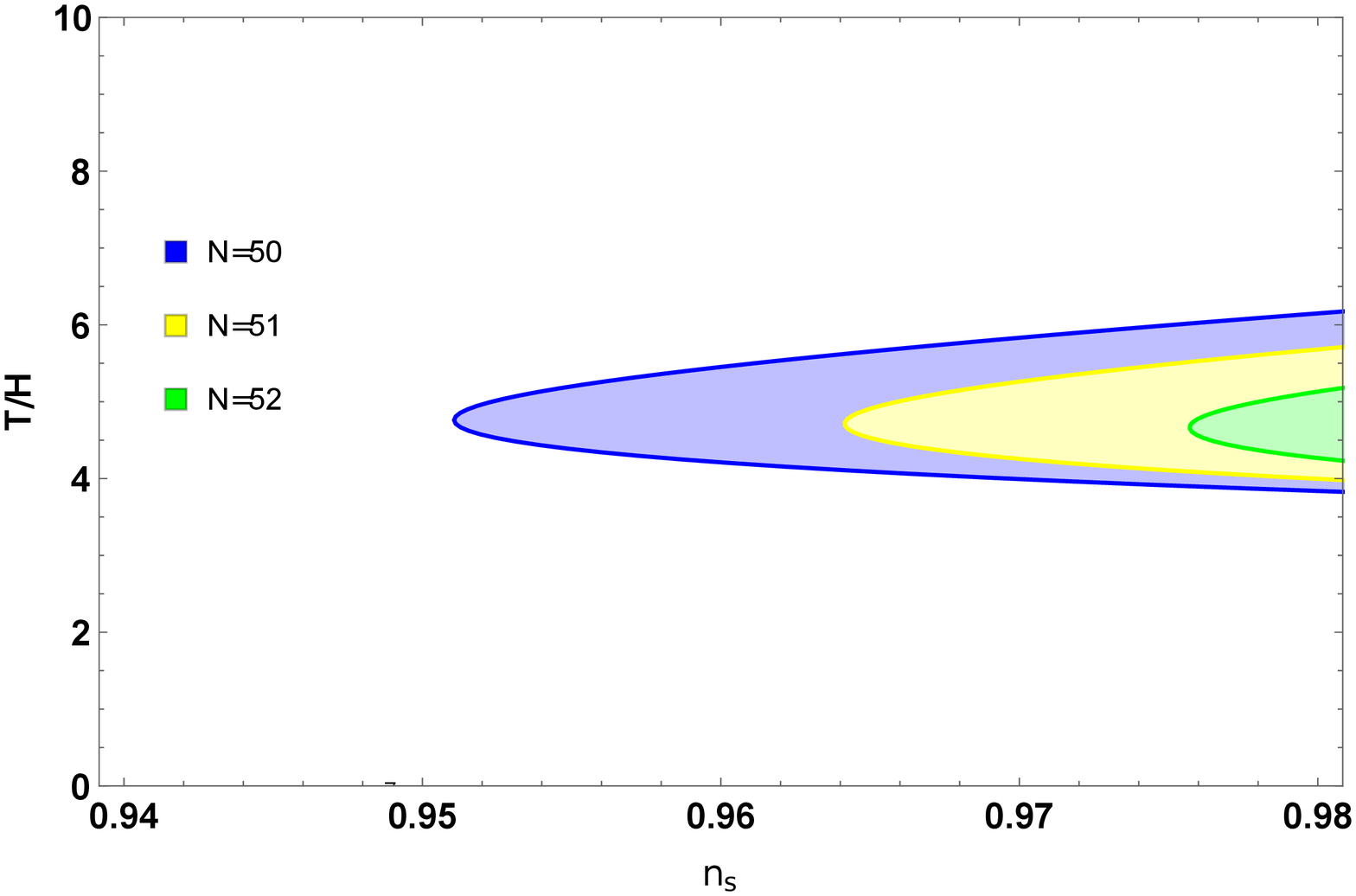}\\
    \end{tabular}
   \caption{The value of $T/H$  versus the spectral index $n_s$ for the conformal anomaly coefficient $c=10^7$
	for $m=-1$,  $C_\phi=10^{-25}$ curve (a) and for $m=3$, $C_\phi=5.7\times 10^{7}$ curve (b).}
\label{THw-1}
\end{figure*}
\begin{figure*}[hbtp]
    \centering
		
\end{figure*}
{
\section{STRONG DISSIPATIVE REGIME $\Upsilon>>3H$ } \label{secVI}

In this section, we analyse the strong dissipative regime, ($Q\gg 1$), by considering exact solutions for the cases $m=3$ and $m\neq 3$. Integrating Eq. (\ref{dotphi1}) by using Eq. (\ref{Hubble parameter}), the evolution of the scalar field  is given by
\begin{equation}\label{phi2}
\phi _{t}-\phi _{0}=\left\{
\begin{array}{cc}
\Big[\frac{(3-m)G_{m}(t)}{2K_{m}}\Big]^{\frac{2}{3-m}}, & \quad \text{for}\;\; m\neq 3 \\
\exp \Big[\frac{G_{m}(t)}{K_{m}}\Big], & \quad \text{for}\;\; m=3
 \end{array}\right.
\end{equation}%
with
\begin{widetext}
\begin{eqnarray}
  G_m(t) &=& H_{A,f,c,m}^s(t)t^{\frac{(8-m)f+2(m-2)}{8}},\label{hyp1}\\
 H_{A,f,c,m}^s(t) &=& {_2F_1}\Big( \frac{m-4}{8},\frac{(8-m)f+2(m-2)}{16(f-1)};1+\frac{(8-m)f+2(m-2)}{16(f-1)};St^{2(f-1)}\Big)\label{hyp2},\\
 K_m &=&\Big(\frac{(8-m)f+2(m-2)}{8}\Big)\Big(\frac{4AfC_{\gamma}}{C_{\phi}}\Big)^{-\frac{1}{2}} \Big(\frac{3 Af(1-f)}{2 \kappa C_{\gamma} }\Big)^{\frac{m-4}{8}}\label{hyp3},
  \end{eqnarray}
 \end{widetext}
where $\phi_{0}$ is an integration constant which will be set to $\phi_{0}=0$ in the rest of the paper, $H_{A,f,c,m}^s(t)$, is a correction term to standard warm inflation and $S=\frac{3A^2f^2}{2 \kappa \rho_c }$.

Since $Q=\Upsilon/3H\gg 1$ and using Eqs. (\ref{Gamma}) and (\ref{Hubble parameter}), the dissipation coefficient $\Upsilon$ can be rewritten in terms of the scalar field as
\begin{equation}\label{Gamma4}
\Upsilon=C_{\phi}  \Big[\frac{-3 Af(f-1)}{2 \kappa C_{\gamma} }\Big]^{\frac{m}{4}} t^{\frac{m}{4}(f-2)}   \phi^{1-m}\Big(1-St^{2(f-1)}\Big)^{\frac{m}{4}}.
\end{equation}
If we compare the amplitude of the scalar inflaton field  Eq (\ref{phi2}) and the dissipation coefficient Eq. (\ref{Gamma4}) at the horizon crossing and for $m\neq3$ with that obtained in the case of standard warm inflation  (see Ref \cite {RamonHerrera}) we find respectively that
\begin{equation}\label{phitks}
\phi_{t_{k}}=[\phi_{t_{k}}]_{\textrm{std}} \Big(H_{A,f,c,m}^s(N)\Big)^{\frac{2}{3-m}},
\end{equation}
and
\begin{equation}\label{Gamma34}
\Upsilon=[\Upsilon]_{\textrm{std}}[\Upsilon]^{s}_{c},
\end{equation}
with
\begin{equation}\label{Gamma34}
[\Upsilon]^{s}_{c}=\Big(H_{A,f,c,m}^s(N)\Big)^{2\frac{(1-m)}{(3-m)}}\Big(1-St_{k}^{2f-2}\Big)^{\frac{m}{4}},
\end{equation}
is the correction term of the dissipation coefficient in the strong dissipation regime.

Fig. \ref{THw-1} shows the plot of $T/H$ versus the spectral index in the weak dissipation regime for $m=-1$ (curve (a)) and $m=3$ (curve (b)). The two curves, (a) and (b), are obtained for the same choice of parameters as in curve (a) and curve (b) in Fig. \ref{figm=3}, respectively. These figures  show clearly that the results obtained in Fig. \ref{figm=3} satisfy the warm condition.

Fig. \ref{gamafort} shows the evolution of the correction term $[\Upsilon]^{s}_{c}$ versus the conformal anomaly coefficient $c$ for three values of the parameter $m$ and for a fixed value of the parameter $f=0.25$ and for three values of the parameter $A$. We notice that the  AdS/CFT correspondence has no significant effect  and the thermal process is described by standard warm inflation for $c<10^7$ for the three chosen values of $m$ parameters. Furthermore, while for $m=0$ and $m = 1$   the thermal process decreases, it increases strongly for $m=-1$ with the increasing values of the conformal anomaly coefficient $c>10^7$   for the range of $f$ and $A$ considered. However this range of the conformal anomaly may change by changing the value of $f$ and $A$.

\begin{figure*}[hbtp]
    \centering
    \begin{tabular}{ccc}
     \includegraphics[width=.45\linewidth]{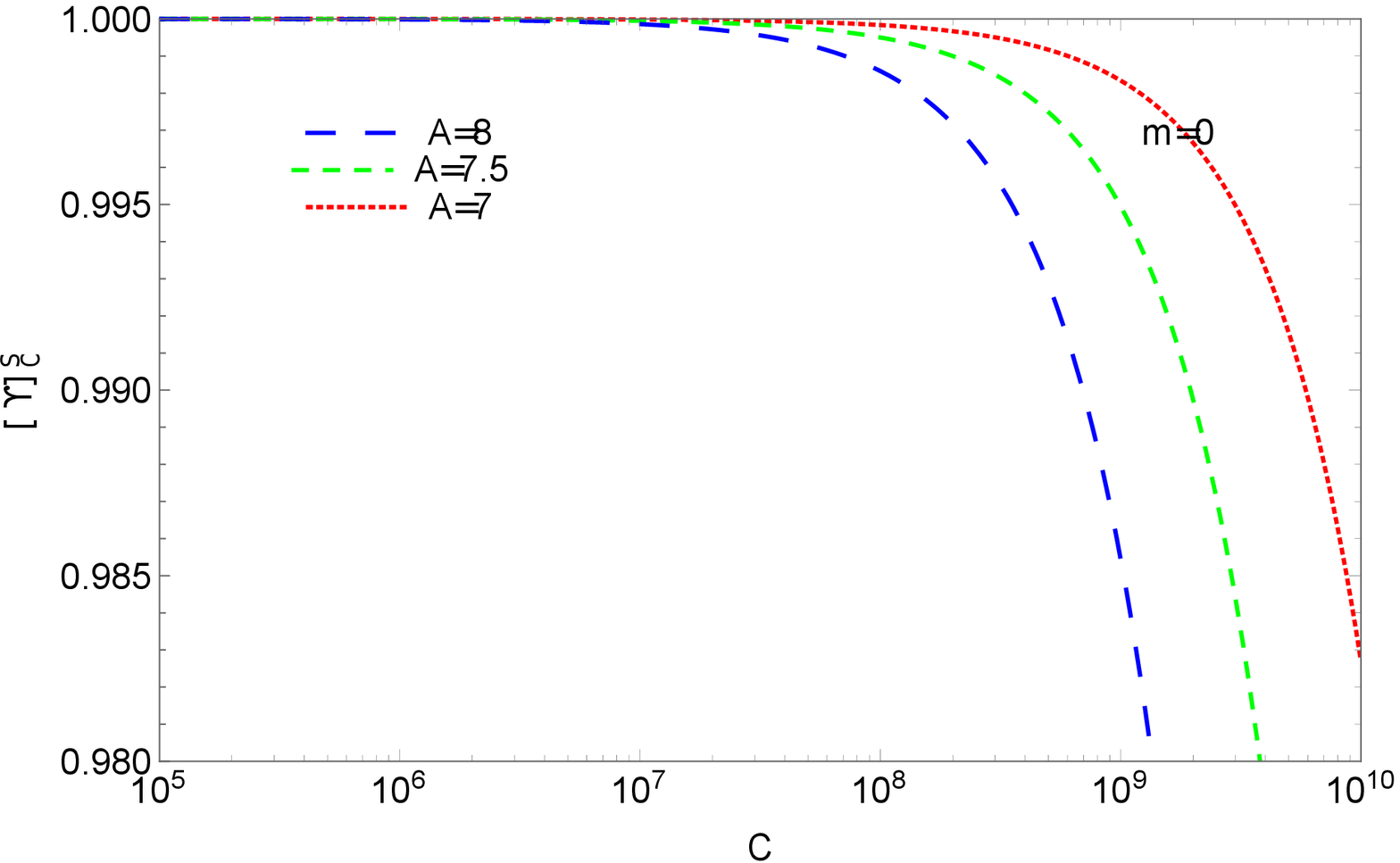} & & \includegraphics[width=.45\linewidth]{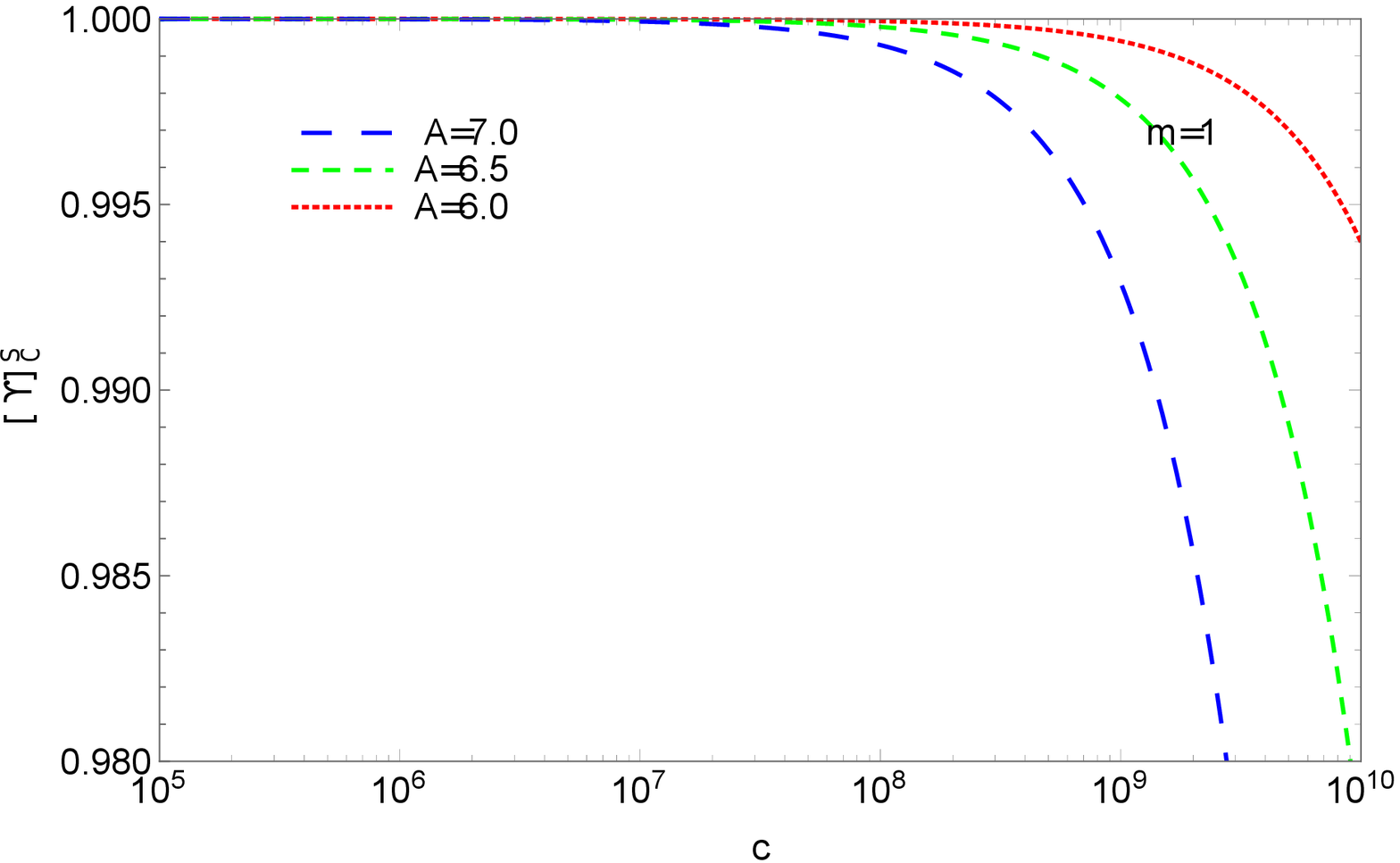}\\
      \end{tabular}
          \includegraphics[width=.45\linewidth]{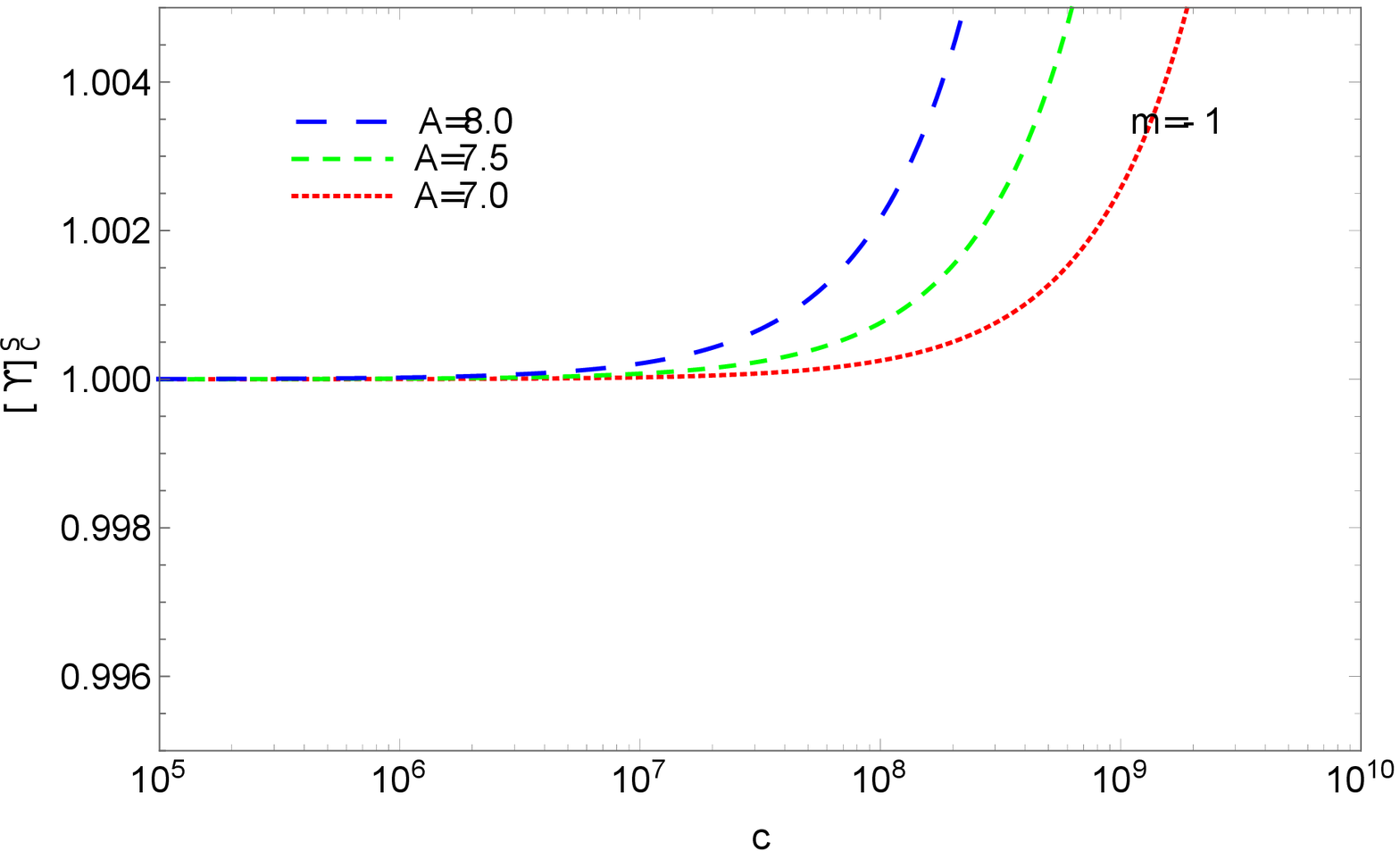} \label{ci}\\

\caption{Evolution of the correction term $[\Upsilon]^s_c$ versus the conformal anomaly coefficient $c$ for $f=0.125$ and for different values of the parameter $A$ for the cases $m=0$ (upper left panel), $m=1$ (upper right panel), and $m=-1$ (lower panel), we have used, $\kappa=1 $, $N=60$ and $C_{\gamma}=70$.}
\label{gamafort}
\end{figure*}
}
{

Furthermore, in the strong dissipative regime,  $1\ll Q$, Eqs. (\ref{PR})-(\ref{R1}) and (\ref{ratio}) take the following form
\begin{widetext}
\begin{eqnarray}
P_{R}&=&\frac{\sqrt{3\pi }A^{2}f^{2}\kappa }{8\pi ^{2}(1-f)}\left[ \frac{%
3Af(1-f)}{2\kappa C_{\gamma }}\right] ^{\frac{1}{4}}I(N)^{\frac{3\left(
3f-2\right) }{4}}(1-S(I(N))^{2f-2})^{-\frac{3}{4}}Q^{\frac{3}{2}}F(Q)\label{PRS}, \\
\frac{T}{H}&=&\frac{1}{Af}\left[ \frac{3Af(1-f)}{2\kappa C_{\gamma }}\right]
^{\frac{1}{4}}I(N)^{\frac{2-3f}{4}}(1-S(I(N))^{2f-2})^{\frac{1}{4}}, \label{THS}\\
Q&=&C_{\phi }\left[ 3Af\right] ^{\frac{4}{4-m}}\left[ \frac{(1-f)}{2\kappa
C_{\gamma }}\right] ^{\frac{m}{4}}I(N)^{f\frac{(m-4)}{4}+\frac{m-2}{2}}\left[
1-S(I(N))^{2f-2}\right] ^{\frac{m}{4}}\phi ^{1-m},\label{RS} \\
r&=&\frac{16(1-f)}{\sqrt{3\pi }}\left[ \frac{3Af(1-f)}{2\kappa C_{\gamma }}%
\right] ^{-\frac{1}{4}}I(N)^{-\frac{\left(f+2\right) }{4},%
}(1-S(I(N))^{2f-2})^{\frac{3}{4}}Q^{-\frac{3}{2}}F(Q)^{-1}.%
\end{eqnarray}
\end{widetext}
\bigskip
We notice here that $\frac{T}{H}$ in Eq. (\ref{THS}) is independent of the parameters $Q$, $m$ and $C_{\phi }$. We notice also that according to the function $F(Q)$ which characterises  the coupling between the inflaton field and radiation, Eqs. (\ref{m3})-(\ref{m-1}), and Eqs. (\ref{PRS}) and (\ref{RS}) that a strong effect is observed on the amplitude of the scalar perturbation in the strong dissipative regime particularly in the case of $m\neq 0$.

As in the weak dissipation case, a numerical approach to calculate the scalar
spectral index  $n_{s}=1-\frac{d\ln (P_{R})}{dN}$ is mandatory.  To this aim, we set the value of the conformal anomaly
coefficient to $c=10^{7}$, we consider as well the warm condition ($\frac{T}{H}>1)$ and the strong dissipation case $(Q\gg 1)$.

The results show that, for  any range of parameters $C_{\phi }$, $A$ and $f$, the dissipative coefficient  $\Upsilon_1$ and $\Upsilon_{0}$ are excluded by observational data.
We plot the dissipative parameters and the tensor-to-scalar ratio ($Q$ and $r$) as parametric functions of the spectral index $n_{s}$. For the dissipative parameter $Q$, Fig. \ref{rs0}, the range of parameters considered are $C_{\phi }=10^{10}$, $0.1<A<6$ and $f=0.0625$. While for the dissipative parameter $Q$, Fig. \ref{rs3}, the range of parameters considered are $C_{\phi }=1.5\times 10^{5}$, $0.1<A<4$, and $f=0.25$.
We notice from the plots of Figs. \ref{rs0} and \ref{rs3}, that we are, indeed, in the strong dissipative regime. For the plots of the same Figs. \ref{rs0} and \ref{rs3}  i.e. $m=1$ and $m=3$ in the dissipative coefficient, we notice that the tensor-to-scalar ratio and the spectral index values agrees with the $1\sigma$ contour of Plank data for a large values of $N$-efolds. In Fig. \ref{THs} we show the evolution of $T/H$ versus the spectral index in the strong regime for $c=10^7$, $0.1<A<6$ and $f=0.25$. The curve  confirms that the result obtained in the strong dissipation regime satisfy the conditions of warm inflation. This means that the strong dissipative regime of warm inflation in the holographic context under consideration is in agreement with the observational data for the choice of parameters mentioned above.
\begin{figure*}[hbtp]
    \centering
    \begin{tabular}{ccc}
		(a)   & \qquad\qquad & (b)  \\
     \includegraphics[width=.45\linewidth]{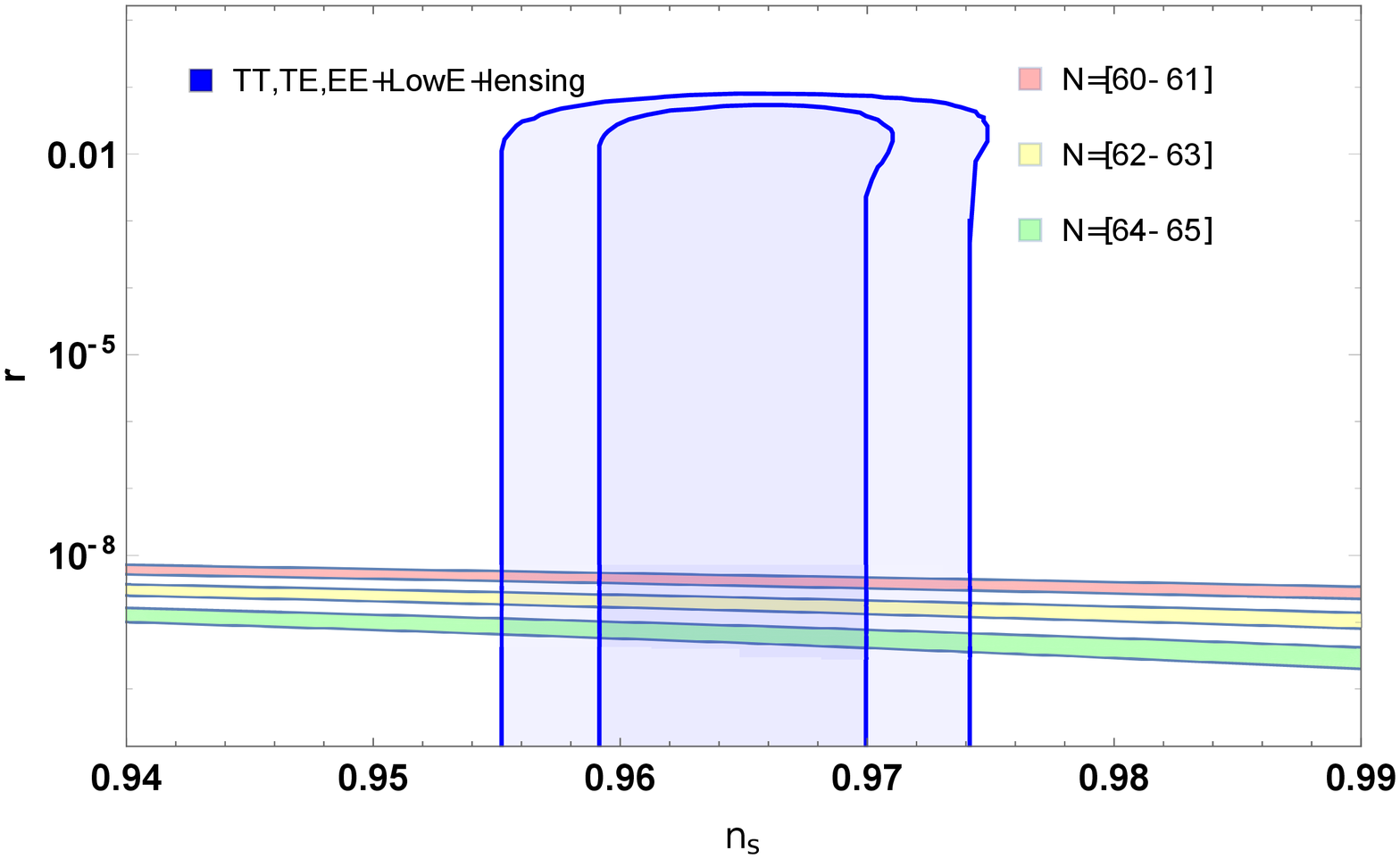} & \qquad\qquad & \includegraphics[width=.45\linewidth]{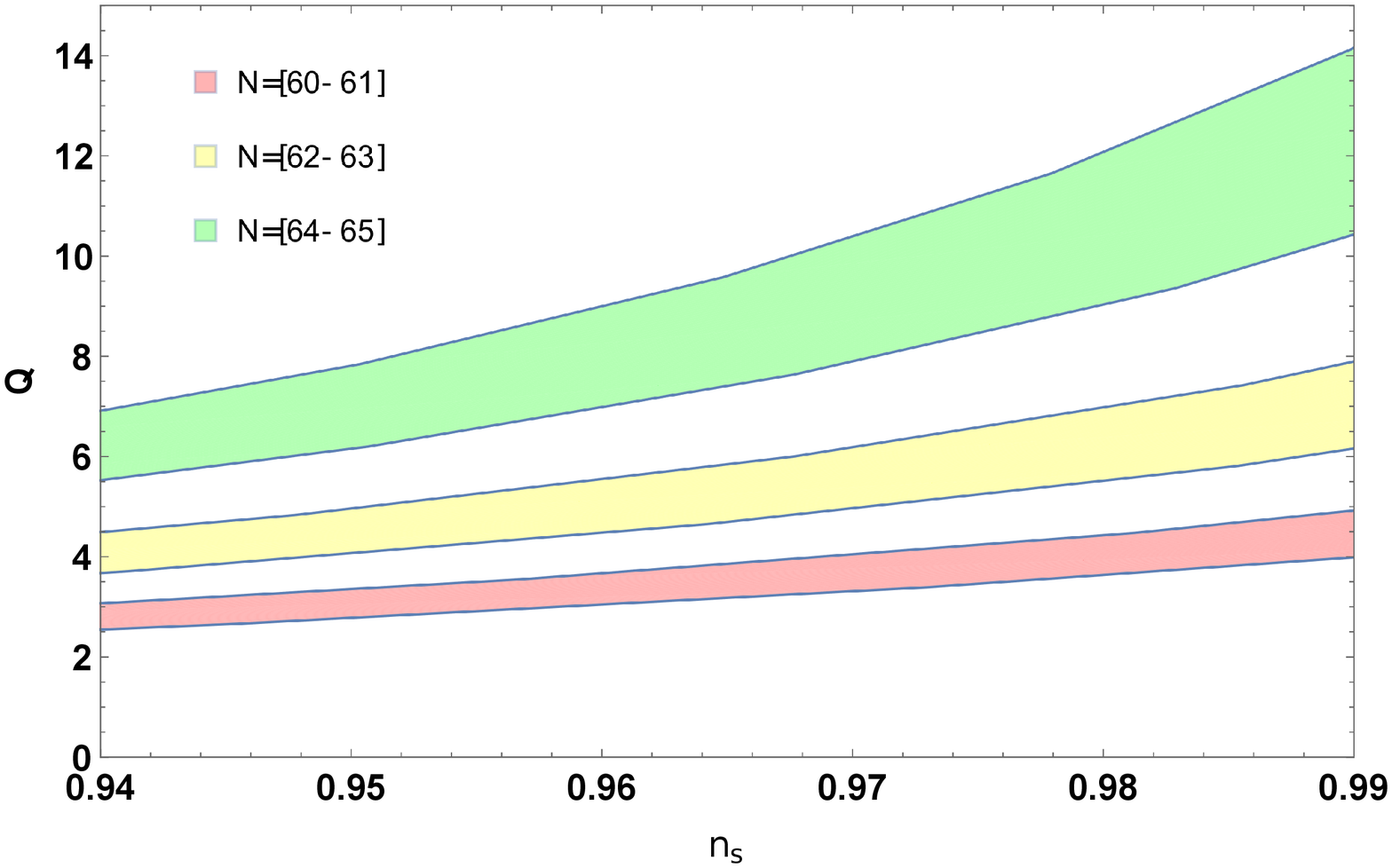}\\
      \end{tabular}
			
  \caption{The  plots show the variation of the dissipative parameter $Q$ (plot (a)) and the variation of the tensor-to-scalar ratio $r$ (plot (b)) versus the spectral index $n_s$ for different values of the e-fold number, the conformal anomaly coefficient $c=10^7$ and the parameter $m=1$. These plots are obtained for  the following choice of parameters  $C_{\phi }=10^{10}$, $0.1<A<6$ and $f=0.0625$.}
\label{rs0}
\end{figure*}
\begin{figure*}[hbtp]
    \centering
    \begin{tabular}{ccc}
		(a)   & \qquad\qquad & (b)  \\
     \includegraphics[width=.45\linewidth]{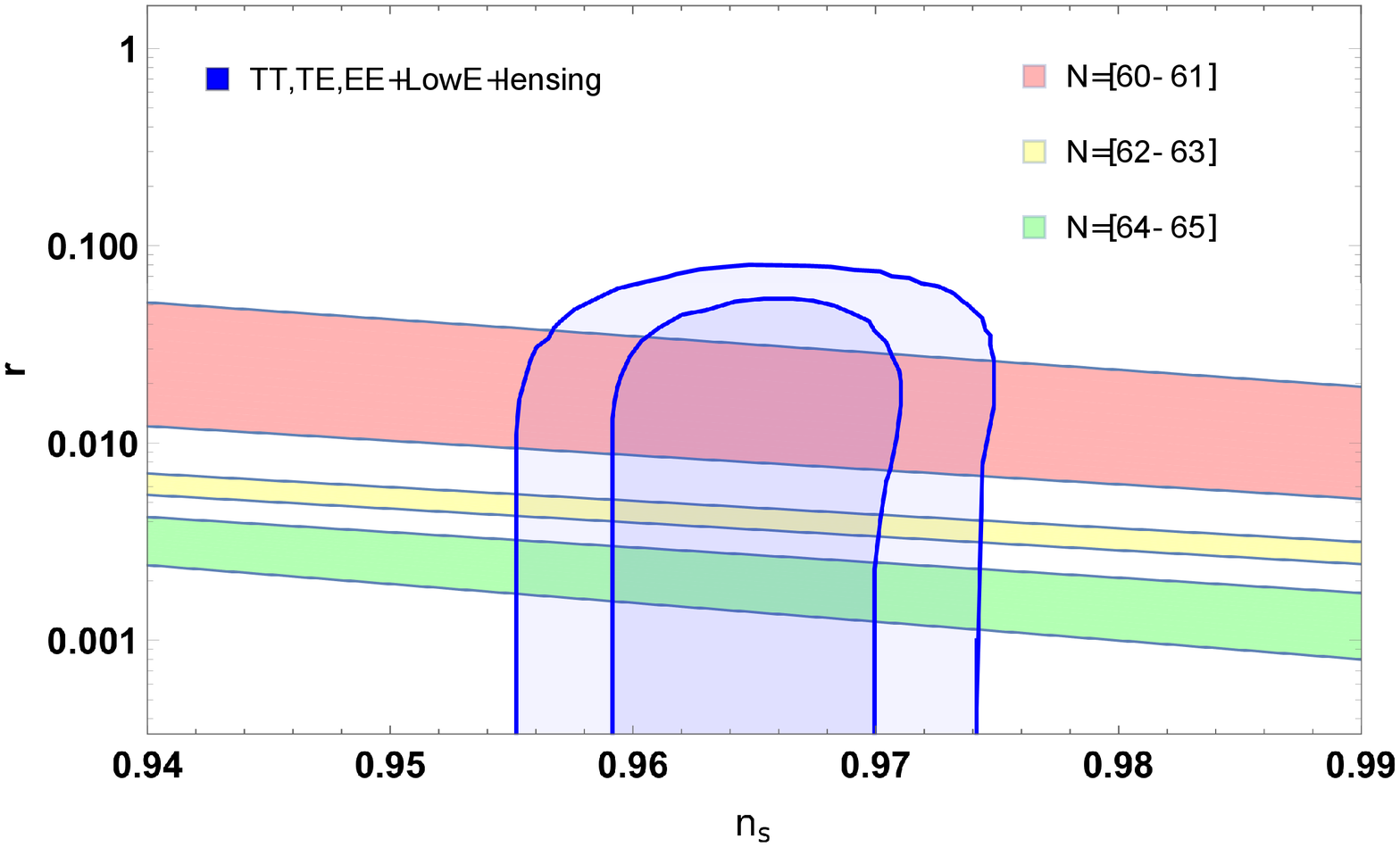} & \qquad\qquad & \includegraphics[width=.45\linewidth]{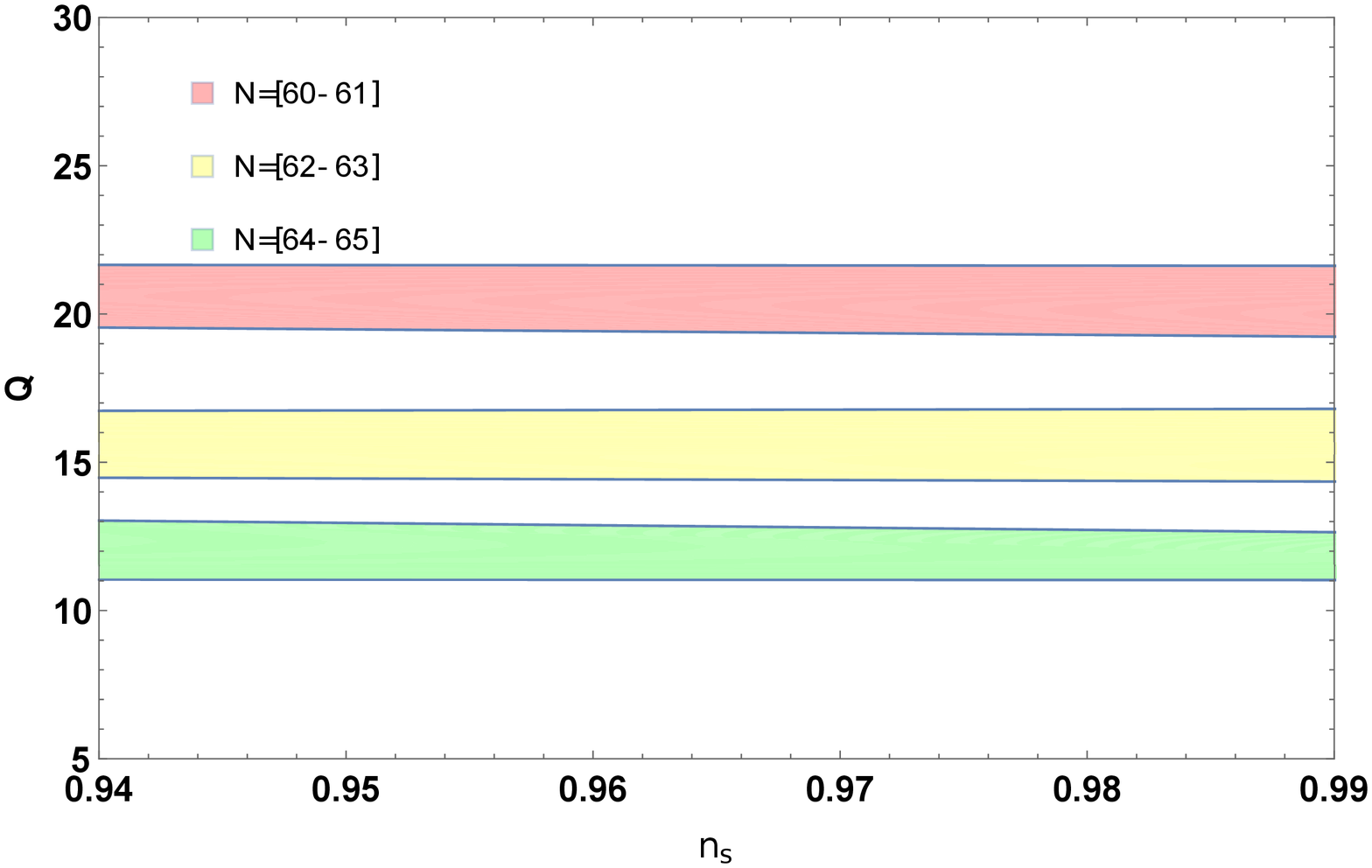}\\
      \end{tabular}
			
  \caption{The  plots show the variation of the dissipative parameter $Q$ (plot (a)) and the variation of the tensor-to-scalar ratio $r$ (plot (b)) versus the spectral index $n_s$ for different values of the e-fold number, the conformal anomaly coefficient $c=10^7$ and the parameter $m=3$. These plots are obtained for  the following choice of parameters  $C_{\phi }=1.5\times 10^{5}$, $0.1<A<4$, and $f=0.25$.}
\label{rs3}
\end{figure*}
\begin{figure*}[hbtp]
    \centering
		\includegraphics[width=.5\linewidth]{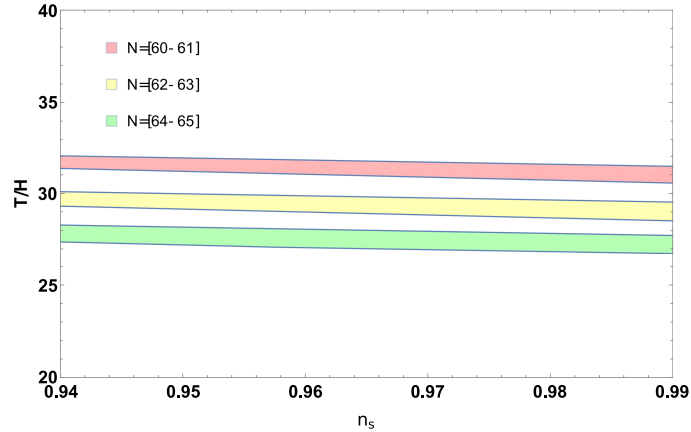}
   \caption{The strong dissipation regime value of $T/H$  versus the spectral index for the conformal anomaly coefficient $c=10^7$, $0.1<A<6$ and $f=0.25$  .}
\label{THs}
\end{figure*}
}
{
\section{Conclusions}\label{secVII}
In this work, we have studied  warm inflation with a scalar field in the context of the AdS/CFT correspondence and with a general form for the dissipative coefficient. We have considered that the scale factor evolve with time during intermediate inflation as $a(t)=a_{I}\exp{At^f}$, ($0<f<1$). In this context, we have provided the basic equation in the weak and the strong dissipation regime and we have shown that these equations are equal to the standard one times some corrections term which at the low energy limit tends to one. We have discussed our model in the frame of warm inflation condition ($T/H>1$) by determining the consistent range of $C_{\phi}$-parameter of the dissipative coefficent, $A$ and $f$ parameters of the intermediate inflation (see Figs. \ref{figm=-1}, \ref{figm=3}, \ref{rs0} and Fig. \ref{rs3}).\\

We have examined our theoretical prediction by plotting the evolution of different inflationary parameters versus observational data.  In both dissipative regimes, while our model is  well supported by Planck data  for the dissipative coefficient $\Upsilon_{3}$, it is not the case for the dissipative coefficient $\Upsilon_{0}$. However, while the dissipative coefficient $\Upsilon_{-1}$ is  well supported by Planck data in the weak dissipative regime, the dissipative coefficient $\Upsilon_{1}$ is well supported   in the strong dissipative regime. \\

We have shown that for a suitable interval of $C_{\phi}$-parameters of the dissipative coefficient and for $A$ and $f$ parameters of the intermediate inflation our predicted inflationary parameters are consistent with the $1\sigma$ confidence level contours derived from Planck data.\\

Finally, we conclude that scalar warm inflation in the context of holographic cosmology for the cases $m=-1$, $m=1$ and $m=3$ may describe the inflationary era and predicts the appropriate inflationary parameters with respect to the observational data within a slow roll approximation for the weak, strong and both regime, respectively.\\

\section{Acknowledgment}\label{sec6}
We thank the referee for her/his suggestion concerning the perturbation issue. The work of M. B. L. is supported by the Basque Foundation of Science Ikerbasque. She also would
like to acknowledge the support from the Basque government Grant No. IT956-16 (Spain) and from
the Grant No. PID2020–114035 GB-100, funded by MCIN/ 638AEI/10.13039/501100011033 and by
“ERDF A way of making Europe.” She is also partially supported by the grant FIS2017-85076-P,
funded by MCIN/ 638 AEI/10.13039/501100011033 and by “ERDF A way of making Europe.''

}
\end{document}